\documentclass{aa}  
\usepackage{graphicx}
\usepackage{amsmath}
\usepackage{amsfonts}
\usepackage{amssymb}
\usepackage{gensymb}
\usepackage{textcomp}
\usepackage[varg]{txfonts}
\usepackage[T1]{fontenc}
\usepackage{natbib}
\usepackage{epstopdf}

\bibpunct{(}{)}{;}{a}{}{,}
\begin{document}

\title{Extreme-ultraviolet bursts and nanoflares in the quiet-Sun transition region and corona}

\author{L.~P.~Chitta\inst{1}, H.~Peter\inst{1}, \and P. R. Young\inst{2,3}}

\institute{Max-Planck-Institut f\"ur Sonnensystemforschung, Justus-von-Liebig-Weg 3, 37077 G\"ottingen, Germany\\
\email{chitta@mps.mpg.de}
\and
NASA Goddard Space Flight Center, Code 671, Greenbelt, MD 20771, USA
\and
Northumbria University, Newcastle Upon Tyne NE1 8ST, UK
}

   \date{Received 23 November 2020 / Accepted 31 January 2021}

\abstract
{The quiet solar corona consists of myriads of loop-like features, with magnetic fields originating from network and internetwork regions on the solar surface. The continuous interaction between these different magnetic patches leads to transient brightenings or bursts that might contribute to the heating of the solar atmosphere. The literature on a variety of such burst phenomena in the solar atmosphere is rich. However, it remains unclear whether such transients, which are mostly observed in the extreme ultraviolet (EUV), play a significant role in atmospheric heating. We revisit the open question of these bursts as a prelude to the new high-resolution EUV imagery expected from the recently launched Solar Orbiter. We use EUV image sequences recorded by the Atmospheric Imaging Assembly (AIA) on board the Solar Dynamics Observatory (SDO) to investigate statistical properties of the bursts. We detect the bursts in the 171\,\AA\ filter images of AIA in an automated way through a pixel-wise analysis by imposing different intensity thresholds. By exploiting the high cadence (12\,s) of the AIA observations, we find that the distribution of lifetimes of these events peaks at about 120\,s. However, a significant number of events also have lifetimes shorter than 60\,s. The sizes of the detected bursts are limited by the spatial resolution, which indicates that a larger number of events might be hidden in the AIA data. We estimate that about 100 new bursts appear per second on the whole Sun. The detected bursts have nanoflare-like energies of $10^{24}$\,erg per event. Based on this, we estimate that at least 100 times more events of a similar nature would be required to account for the energy that is required to heat the corona. When AIA observations are considered alone, the EUV bursts discussed here therefore play no significant role in the coronal heating of the quiet Sun. If the coronal heating of the quiet Sun is mainly bursty, then the high-resolution EUV observations from Solar Orbiter may be able to reduce the deficit in the number of EUV bursts seen with SDO/AIA at least partly by detecting more such events.}   

   \titlerunning{Extreme-ultraviolet bursts and nanoflares in the quiet-Sun transition region and corona}
   \authorrunning{L. P. Chitta et al.}

   \maketitle

\section{Introduction\label{sec:int}}

Coronal emission from hot plasma in the quiet Sun outside active regions is mostly observed in the form of a diffuse background that is loosely related to the magnetic concentrations in the chromospheric network. Impulsive behavior on timescales shorter than typical cooling times of the plasma is also found, however, which is indicative of transient heating of the gas to temperatures of 1\,MK or more. This is best visible in extreme-ultraviolet (EUV) emission. On scales of over 10\,Mm, coronal bright points exhibit flaring activity \citep[][]{1997ApJ...488..499K}, which might subsequently explode as the underlying photospheric magnetic flux converges \citep[][]{1994ApJ...427..459P}, and launch the hot plasma jets that are observed in soft X-ray images \citep[][]{2018ApJ...859....3M,2019LRSP...16....2M}. These explosive events are indeed common throughout different layers of the quiet-Sun atmosphere. The well-studied transition region explosive events that were first observed in HRTS\footnote{The High Resolution Telescope and Spectrograph.} data are most widely known \citep[][]{1983ApJ...272..329B,1989SoPh..123...41D,1994AdSpR..14d..13D}. These events are typically characterized by their spectroscopic features, such as broad emission line profiles or separate components in the blue and red wing (from ion species such as C\,{\sc iv} and Si\,{\sc iv, } which form at about 0.1\,MK), which are indicative of bidirectional plasma jets originating from magnetic reconnection \citep[][]{1997Natur.386..811I}. These transition region explosive events have a spatial scale of 1\,Mm to 2\,Mm, and it is estimated that there are about 600 new events per second on the Sun \citep[][]{1994AdSpR..14d..13D}, with no clear signatures in coronal spectroscopic diagnostics \citep[][]{2002A&A...392..309T}. 

In addition to flaring coronal bright points and transition region explosive events, a variety of quiet-Sun transient events have been documented in the literature. Blinkers \citep[][]{1997SoPh..175..467H}, EUV transients \citep[][]{1998A&A...336.1039B}, and micro- and nanoflares or microevents \citep[][]{ 1998ApJ...501L.213K,2000ApJ...529..554P,2000ApJ...535.1027A,2000ApJ...535.1047A,2002ApJ...568..413B} have been extensively studied using SOHO\footnote{The Solar and Heliospheric Observatory.} and TRACE\footnote{The Transition Region and Coronal Explorer.} observations. This diverse classification might arise because different diagnostics (spectroscopic vs. imagery), different instruments, and different methods of event detection were employed \citep[][]{2003A&A...409..755H}. Many of these different classes might be identical or at least based on the same physical processes.

\begin{figure*}
\begin{center}
\includegraphics[width=0.49\textwidth]{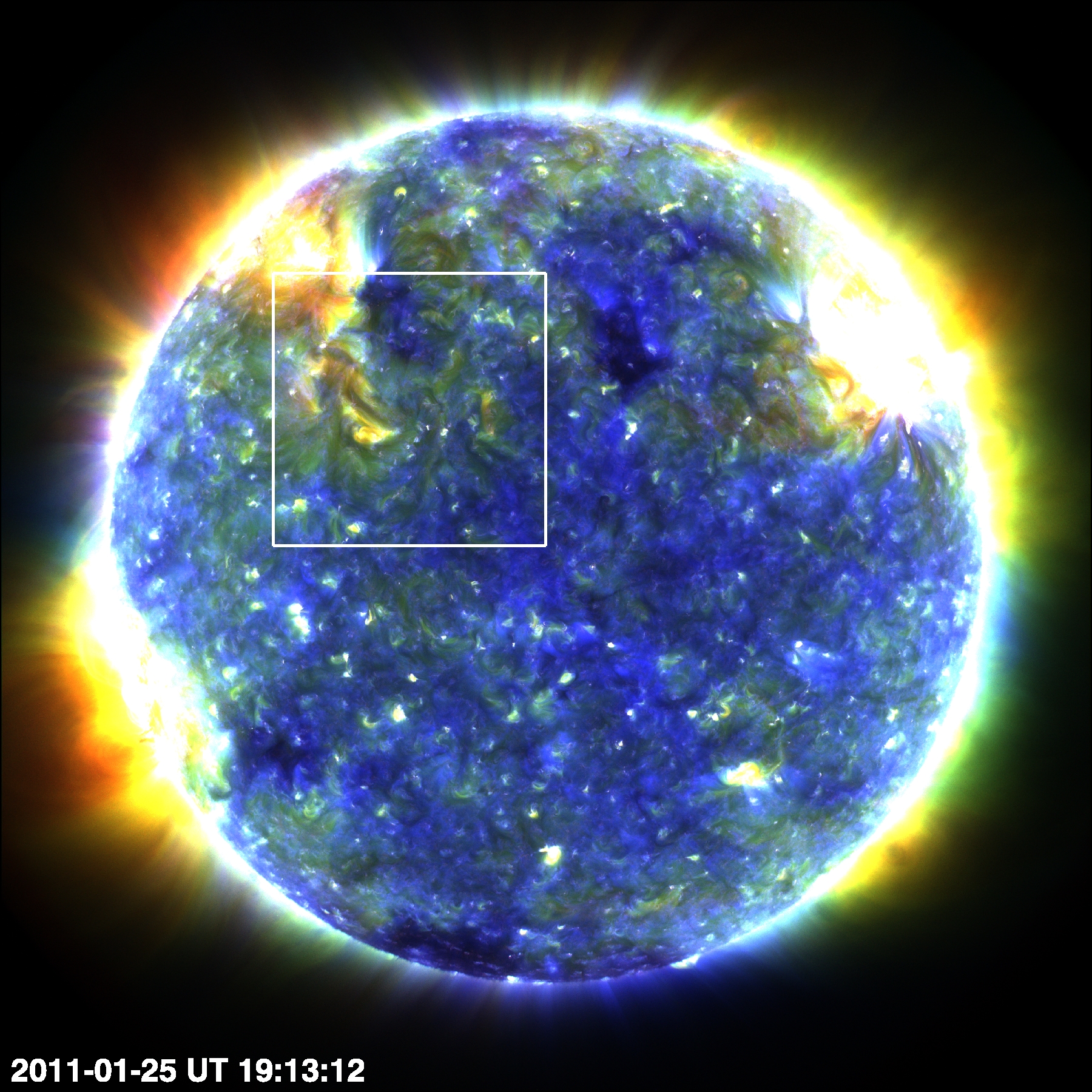}
\includegraphics[width=0.49\textwidth]{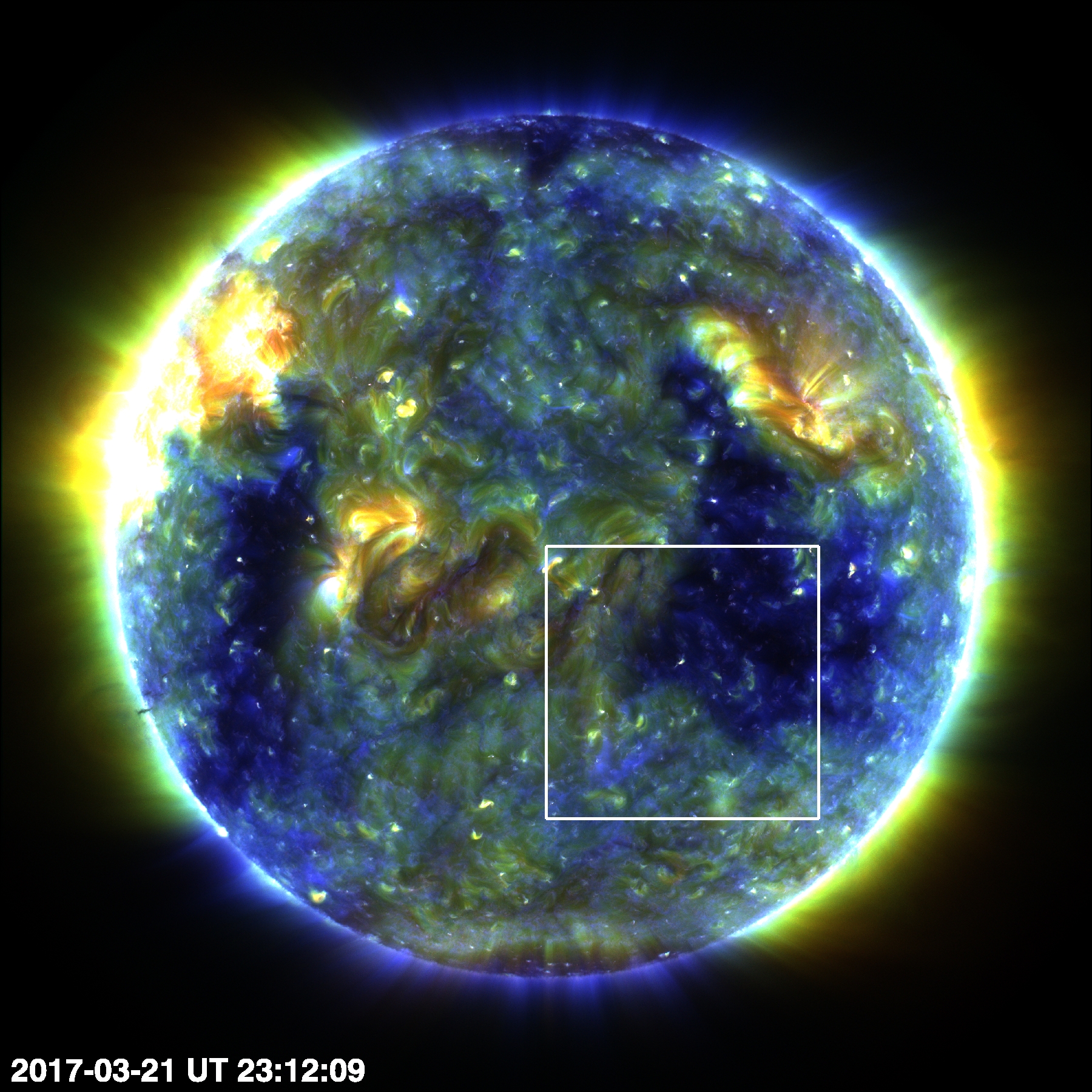}
\caption{Solar corona. Full-disk composite maps of the solar corona observed with the three EUV filters on SDO/AIA. In both panels, the shaded blue, green, and red regions represent plasma emission detected with the 171\,\AA\ (maximum 250\,DN\,s$^{-1}$), 193\,\AA\ (maximum 250\,DN\,s$^{-1}$), and 211\,\AA\ (maximum 75\,DN\,s$^{-1}$) filters, respectively. The white boxes in both panels mark the quiet-Sun regions we analyzed (the field of view is $614\arcsec\times 614\arcsec$). Solar north is up. See Sects.\,\ref{sec:obs} for details.\label{fig:fd}}
\end{center}
\end{figure*}

Based on event occurrence rates and energy distributions from these observations, \citet{2000ApJ...535.1047A} argued that micro- and nanoflares are insufficient to balance the quiet-Sun energy losses. These earlier observations of coronal brightenings were limited by either spatial resolution or cadence of observations, however. For example, although TRACE had a high angular resolution of about 1\arcsec, the datasets used to detect micro- and nanoflares had a cadence of only 100\,s or longer \citep[see][]{2000ApJ...535.1027A,2000ApJ...529..554P}. The SOHO Extreme ultraviolet Imaging Telescope observations used to detect transients had a moderate angular resolution of 5\arcsec\ with a cadence of 60\,s. These observations would have missed  smaller events at shorter timescales similar to transition region explosive events with average lifetimes of 60\,s \cite[][]{1994AdSpR..14d..13D}.

The EUV images observed with the Atmospheric Imaging Assembly \citep[AIA;][]{2012SoPh..275...17L} onboard the Solar Dynamics Observatory \citep[SDO;][]{2012SoPh..275....3P} have a spatial resolution of about 1.4\arcsec\ and are recorded at a higher cadence of 12\,s. The AIA EUV observations are therefore better suited for investigating small-scale quiet-Sun coronal brightenings and bursts that might evolve on timescales of about 60\,s. Recently, \citet{2016A&A...591A.148J} used SDO/AIA data to investigate the properties of coronal brightenings. However, they used 120\,s cadence data (comparable to that of the TRACE observations). This resulted in the detection of events with a duration of at least 240\,s \citep[Fig.\,5 in][]{2016A&A...591A.148J}. 

\begin{figure*}
\begin{center}
\includegraphics[width=0.49\textwidth]{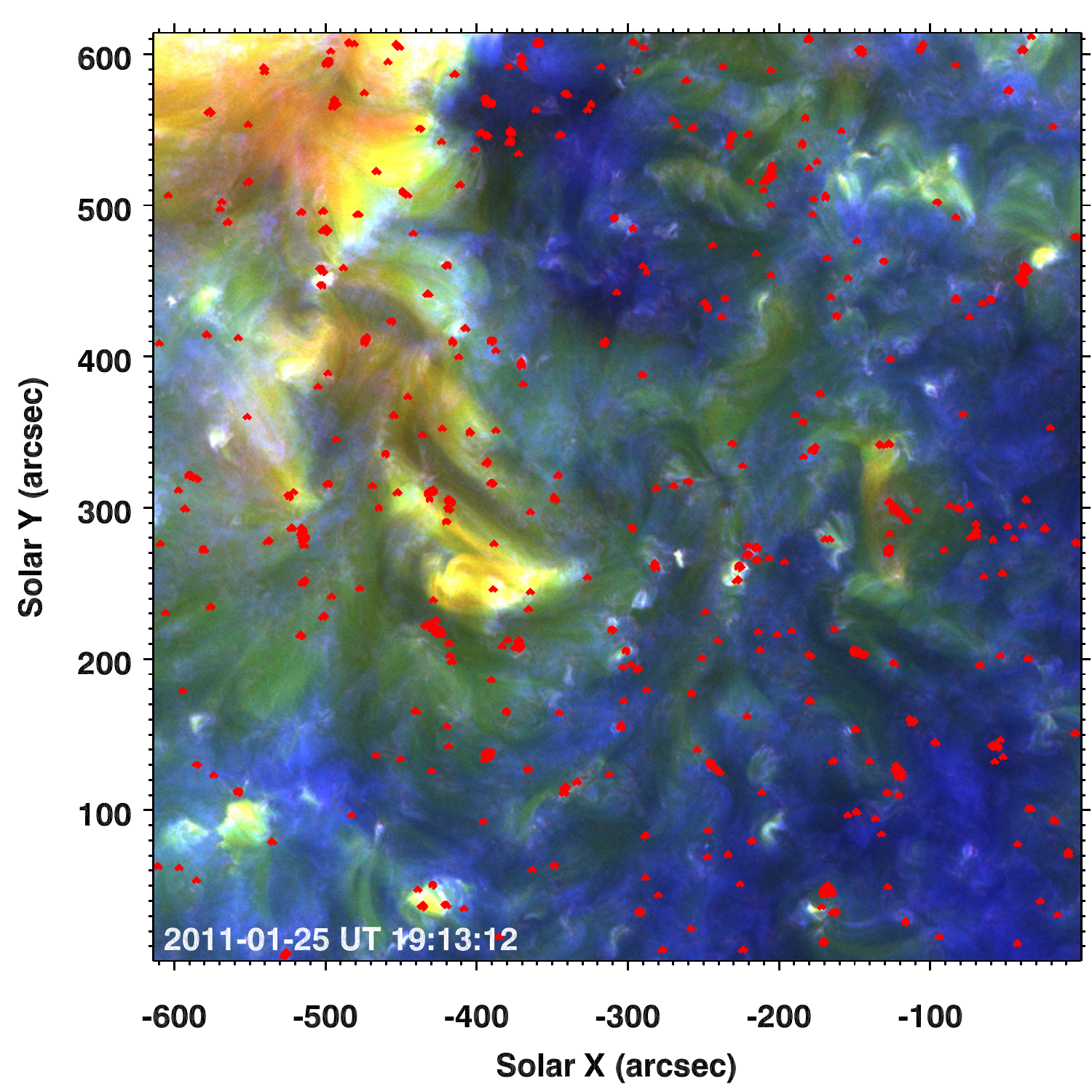}
\includegraphics[width=0.49\textwidth]{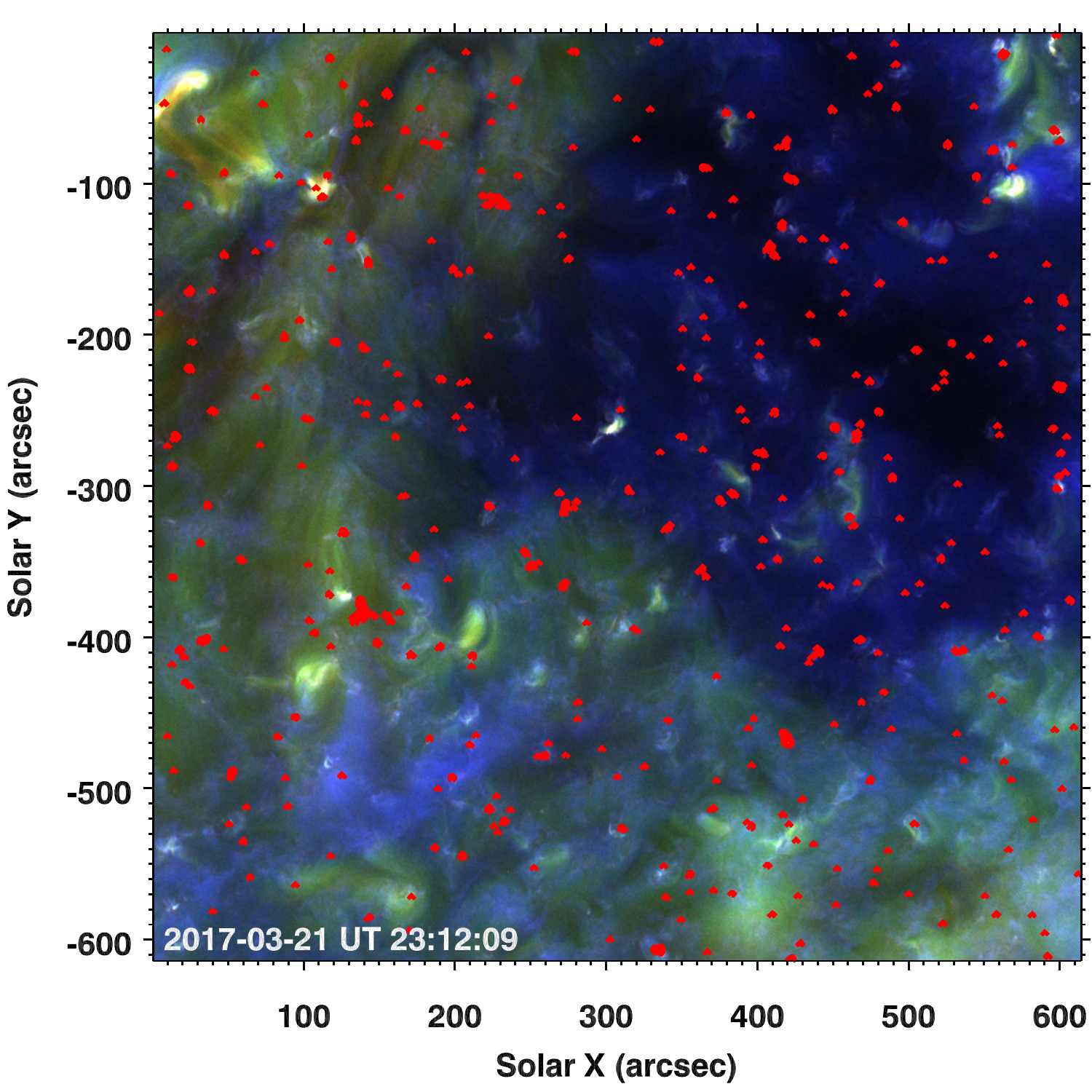}
\includegraphics[width=0.49\textwidth]{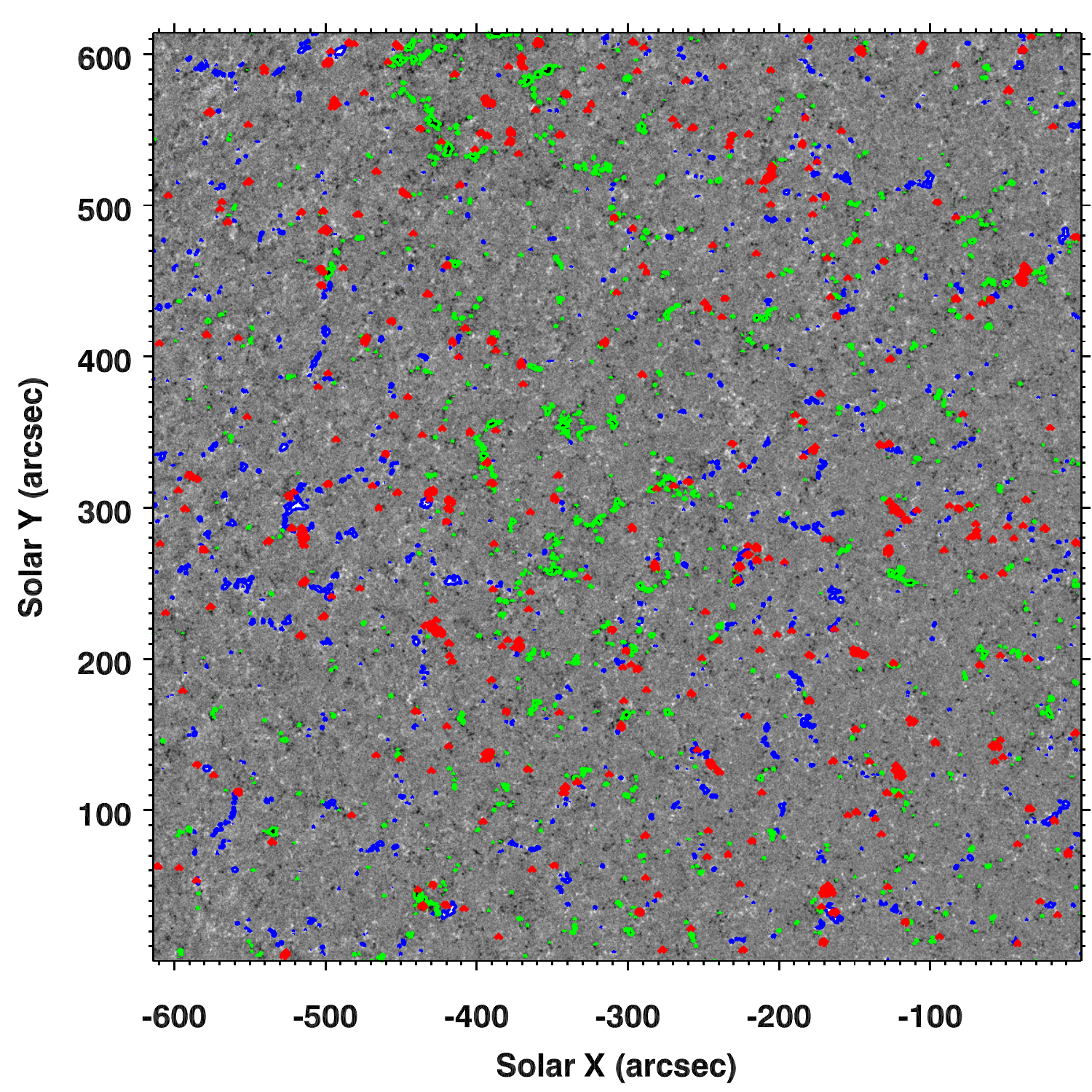}
\includegraphics[width=0.49\textwidth]{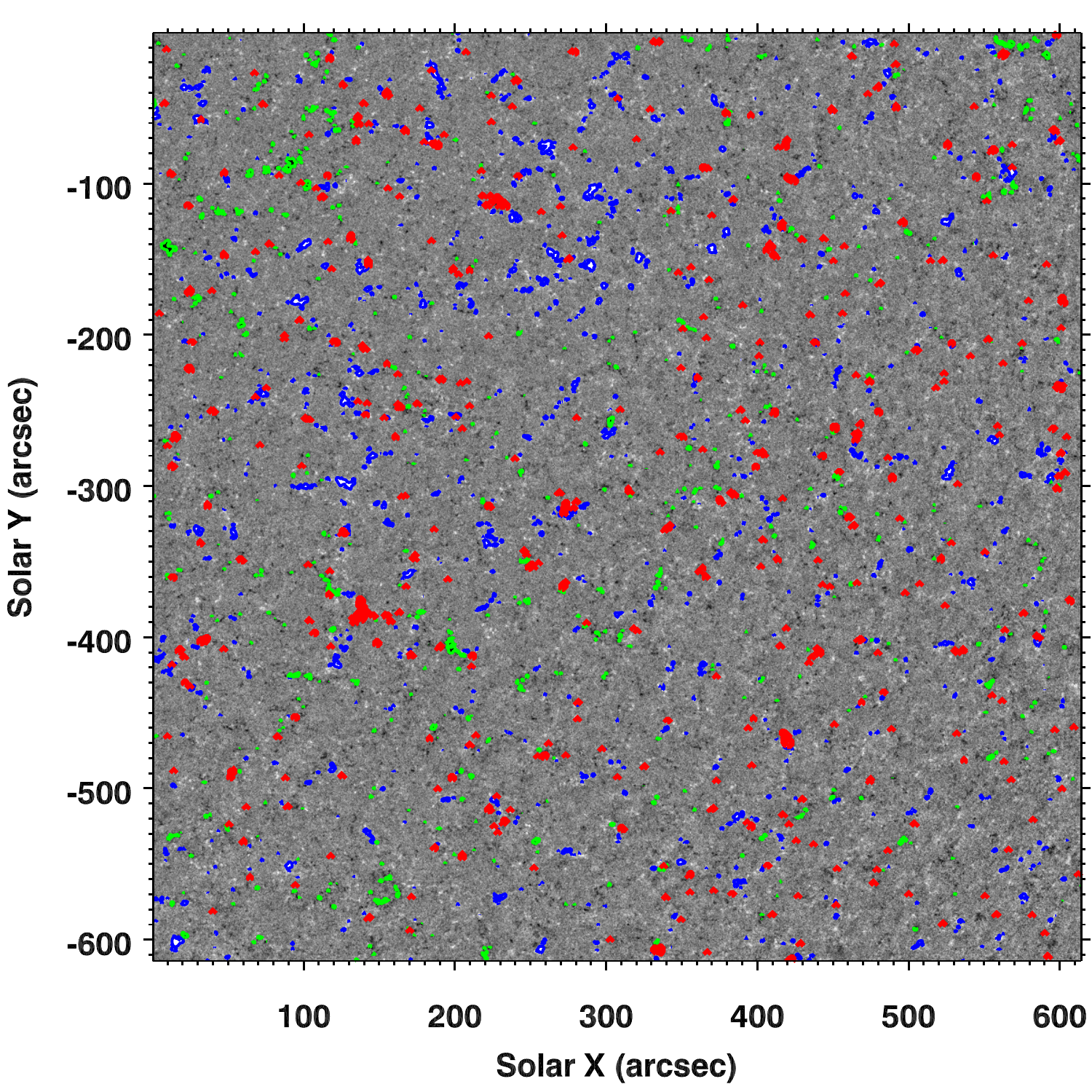}
\caption{Overview of the quiet-Sun coronal regions we analyzed and the underlying surface magnetic fields. A closer view of the coronal regions, outlined by white boxes in Fig.\,\ref{fig:fd}, is presented in the top panels. The field of view is $614\arcsec\times 614\arcsec$. The red contours outline spatial locations of EUV bursts in the respective panels at that instance. The bottom panels show the line-of-sight magnetic field maps observed with SDO/HMI in grayscale. They are cotemporal with respect to the top panels, which are overlaid with red contours. The maps are saturated at $\pm50$\,G. The dark shaded regions represent south or negative-polarity magnetic field patches, and the light shaded regions represent north or positive-polarity magnetic field patches. The blue and green contours cover positive- and negative-polarity magnetic patches, respectively, with flux densities above 50\,G. See Sects.\,\ref{sec:obs}, \ref{sec:det}, and \ref{sec:prop} for details.\label{fig:region}}
\end{center}
\end{figure*}

In this study, we revisit EUV bursts and take advantage of the 12\,s\ high-cadence EUV imagery produced by the SDO/AIA. Our emphasis lies on the properties of small-scale short-duration EUV bursts that have likely been missed by earlier studies. The event statistics we present might be of interest to gain further insights into EUV bursts that will be observed with the recently launched Solar Orbiter \citep[][]{2020A&A...642A...1M}.

\section{Observations and data processing\label{sec:obs}}

To study the statistical properties of EUV bursts in the quiet-Sun corona, we used data recorded in six EUV filters  of SDO/AIA (94\,\AA, 131\,\AA, 171\,\AA, 193\,\AA, 211\,\AA, and 335\,\AA). Furthermore, to verify the consistency of our results, we used two datasets. The first dataset is a 30-minute time sequence of EUV images recorded in the six EUV filters, starting at 19:00\,UT on 2011 January 25. The second dataset is also a 30-minute time sequence of EUV images obtained with these filters, starting at 23:00\,UT on 2017 March 21. 

We obtained the full-disk level-1 SDO/AIA data from the Joint Science Operations Center\footnote{\url{http://jsoc.stanford.edu/}}. These default level-1 data are among other calibration steps corrected for cosmic-ray hits that manifest themselves as sharp local intensity enhancements and are mostly confined to isolated pixels on the image \citep[][]{2012SoPh..275...41B}. Because our focus is on EUV bursts, which may also exhibit sharp intensity enhancement in space and time, we re-spiked the level-1 AIA data using the \texttt{aia\_respike} procedure that is available in the SDO Solarsoft library. Next, these images were coregistered and normalized to their exposure times with the \texttt{aia\_prep} procedure. These AIA image sequences have a time cadence of 12\,s and a spatial sampling of $0.6\arcsec\,\rm{pixel}^{-1}$. Representative full-disk coronal images from these datasets are displayed in Fig.\,\ref{fig:fd}.

The images were then tracked to remove solar rotation. There are small nearly constant spatial offsets (smaller than 1\arcsec) across different AIA filters in the resulting data. To correct for these nearly constant cross-filter offsets, we first extracted large cutouts ($2048\,\rm{pixels}\times 2048\,\rm{pixels}$; centered at the disk center) from each of the six AIA EUV filters from the respective snapshots that are cotemporal with the images displayed in Fig.\,\ref{fig:fd}. We then determined the cross-filter spatial offsets with respect to the 171\,\AA\ image using a cross-correlation technique. These shifts (different for each filter and independent of time) were then applied to the time sequences of the respective filters. This means that each filter time sequence was shifted by only a constant amount with respect to the 171\,\AA\ image sequence, ignoring any minor time-dependence (this is justified because the cross-filter alignment prior to cross-correlation is already better than the spatial resolution of AIA).\footnote{Whenever interpolation is required to better reconstruct the input image, a cubic convolution method (with its parameter set to -0.5) was employed.} To aid discussion, we call these calibrated and processed AIA image sequences the level-2 data product. 

In the next step, we cleaned the level-2 data by removing sharp intensity peaks that were identified in the temporal domain. We achieved this by conducting a pixel-by-pixel analysis in two steps separately for each of the six AIA filter level-2 image sequences. In the first step, we identified peaks in intensity for a given filter that exceeded the mean intensity computed from the time series at each pixel by more than 5$\sigma$. We then compared the intensity of each of the identified peaks with that from the preceding and subsequent snapshots in the time series. When the two intensities next in time to the peak intensity did not exceed the mean intensity by more than 2$\sigma$, the peak is quite sharp and spike-like. This means that the identified spike is very likely caused by a cosmic-ray hit. In this case, we replaced the peak intensity with the average of the intensities from next-in-time steps at that pixel. In the second step, we repeated this process on this newly de-spiked time series by lowering the peak intensity threshold. We recalculated the mean intensity at a given pixel. Then we replaced all intensity peaks that exceeded the newly calculated mean intensity by 4$\sigma$ with the average of the intensities from next-in-time steps at that pixel if they did not exceed the mean intensity by 2$\sigma$. This double de-spiking removed all sharp intensity spikes in the time series that are likely due to cosmic rays. From these cleaned images we considered the time series of a large enough field of view ($1024\,\rm{pixels}\times 1024\,\rm{pixels} \approx 614\arcsec\times 614\arcsec$) for further analysis. With respect to the solar disk center at $(0\arcsec,0\arcsec)$, the field of view of the 2011 dataset is centered at  $(-307.2\arcsec,307.2\arcsec)$ and the 2017 data are centered at $(307.2\arcsec,-307.2\arcsec)$ (the regions marked by the white boxes in Fig.\,\ref{fig:fd}). In Sect.\,\ref{sec:det} we describe the method we used to identify EUV bursts.

To qualitatively investigate the properties of the magnetic fields at the solar surface that underlie the EUV bursts, we complemented AIA observations with the line-of-sight magnetic field maps obtained with the Helioseismic and Magnetic Imager \citep[HMI;][]{2012SoPh..275..207S} on board the SDO. These were also processed with the \texttt{aia\_prep} procedure. These processed HMI data have a time cadence of 45\,s and an image scale of $0.6\arcsec\,\rm{pixel}^{-1}$. 

The time period of the 2017 dataset overlaps the UV observations of the chromosphere and transition region that are recorded with the Interface Region Imaging Spectrograph \citep[IRIS;][]{2014SoPh..289.2733D}. IRIS observed a field of view of about $33\arcsec\times 130\arcsec$ centered at $(431\arcsec,-453\arcsec)$ on the solar disk in a 96-step dense-raster mode between 2017 March 21 23:02\,UT and 2017 March 22 02:26\,UT. The raster step in the scan direction is 0.35\arcsec, with a 2-pixel spatial binning and sampling of 0.33\arcsec along the slit. The exposure time per slit position is 30\,s. The spectral resolution in the far-UV is about 25\,m\AA. We used level-2 IRIS data\footnote{Available at \url{https://iris.lmsal.com/}.} to examine the spectral properties of EUV bursts, in particular with the Si\,{\sc iv} 1394\,\AA\ line, which forms in the transition region at temperatures of about 80\,kK. 

To give an overview of the data, in Fig.\,\ref{fig:region} we display sample AIA and HMI maps of the two datasets we analyzed.  Although a large fraction of the analyzed field of view is covered by a coronal hole (northwest quadrant of the top right panel in Fig.\,\ref{fig:region}) in the 2017 dataset, we did not distinguish it from a quiet-Sun corona in the remaining analysis (see the discussion at the end of Sect.\,\ref{sec:prop}).   

\begin{figure}
\begin{center}
\includegraphics[width=0.49\textwidth]{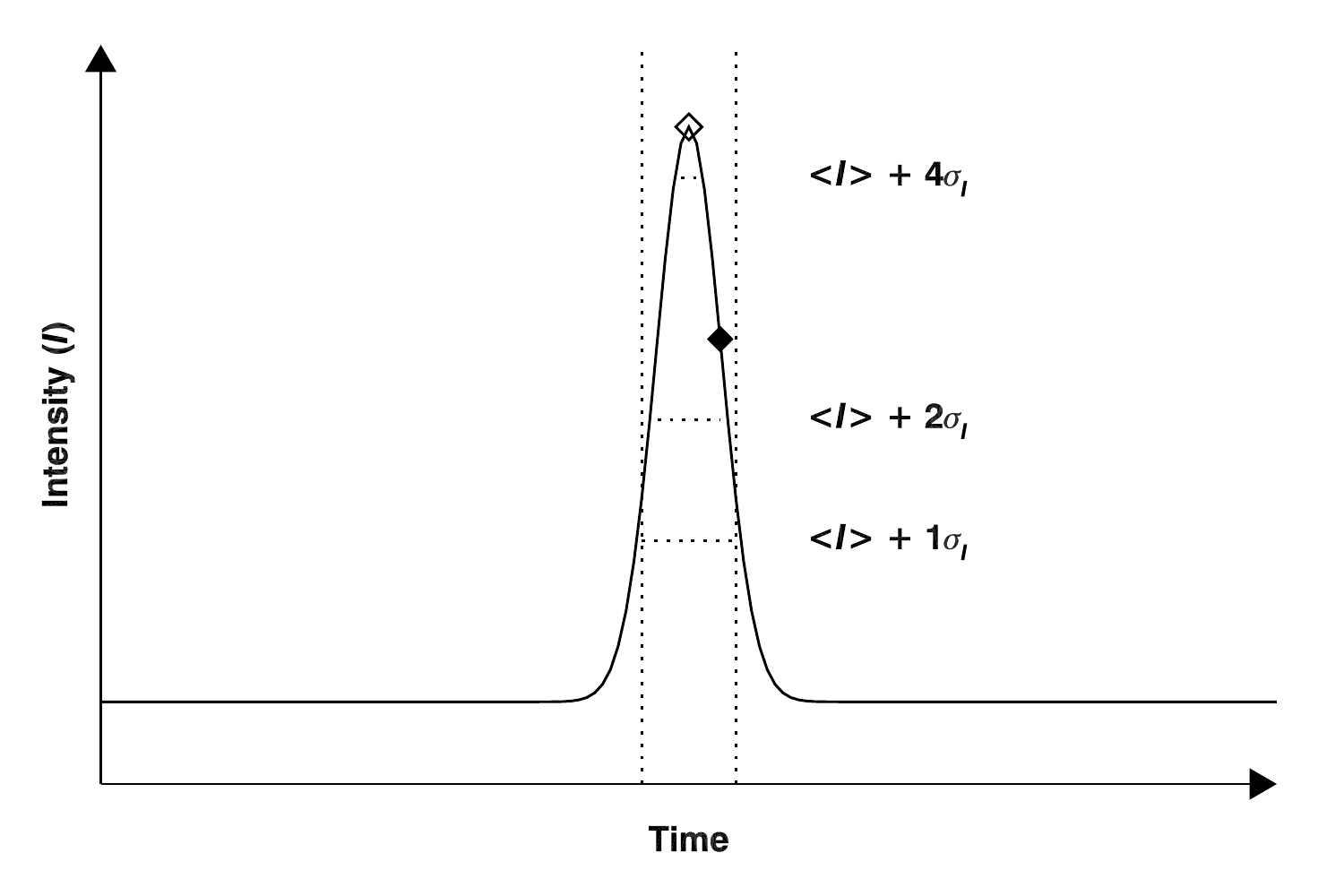}
\caption{Illustration of a detected EUV burst. The solid curve represents the AIA 171\,\AA\ intensity, $I$, as a function of time at a given pixel in the image sequence. The horizontal dotted lines from top to bottom are four, two, and one times the standard deviation of the light curve ($\sigma_I$) above the mean intensity, $<I>$. The open diamond marks the peak that exceeds $<I> +\ 4\sigma_I$. In this case, the intensity at the next time step after the peak is greater than  $<I> +\ 2\sigma_I$. The light curve is therefore flagged as a valid EUV burst signal according to our criteria in Sect.\,\ref{sec:det}. Following the light curve forward and backward in time from the peak, the lifetime is defined as the duration between two instances when the signal is still greater than $<I> +\ 1\sigma_I$ (i.e., the length of time that is bounded by the two vertical dotted lines).
\label{fig:illus}}
\end{center}
\end{figure}

\section{EUV bursts in the quiet-Sun corona\label{sec:burst}}
EUV bursts appear as compact small-scale (a few arcsec) brightenings in the quiet-Sun corona. While some of the bursts can be identified visually, we resorted to an automatic detection of these features in the time series by imposing intensity thresholds to retrieve their statistical properties
(e.g., lifetime, occupied area, and number of events). We describe our method of detecting EUV bursts and present the main findings. 

\subsection{Detection\label{sec:det}}
We used the clean twice de-spiked level-2 AIA 171\,\AA\ image sequence data product (see Sect.\,\ref{sec:obs} for details) to identify the EUV bursts. In our definition, EUV bursts are intensity enhancements above a certain imposed threshold in the temporal domain. We first identified peaks in the time series at each pixel. We set the threshold to be 4$\sigma$ above the mean intensity of the time series at that pixel. The second of the de-spiking steps described in Sect.\,\ref{sec:obs} replaces any sharp intensity spike in the time series above this threshold with the average intensity from adjacent time steps. The newly identified peaks above the threshold are therefore not as sharp. Furthermore, to ensure that the identified peaks are of solar origin, we imposed another condition: the EUV intensity recorded during at least one of the adjacent time steps (either preceding or following peak) must be greater than 2$\sigma$ above the mean intensity of the time series at that pixel. Unlike sharp peaks caused by cosmic-ray hits, this condition means we identify only those intensity peaks that are temporally resolved in two or more exposures in the time series by exhibiting either gradual increase and or decrease in intensity. Moreover, because the duration of the time series is finite, there could be boundary effects when we follow the evolution of EUV bursts that occur close to the beginning or end of the time series. To mitigate this, we did not take any peaks into account that occurred within 2\,minutes from the start or end of the time series. With a cadence of 12\,s, this corresponds to ten AIA snapshots. This pixel-by-pixel analysis of the burst identification also ensures that any persistent bright regions (e.g., coronal bright points) are excluded, unless they exhibit 4$\sigma$ intensity enhancements above the local mean. For a given valid 4$\sigma$ peak, we then analyzed the intensity of the adjacent eight pixels. When the intensity of any of the pixels that were connected to the central peak exceeded the mean at that pixel by 3$\sigma$, but remained lower than 4$\sigma$ at the same time as the previously described valid peak, we also flagged this connected pixel to be a part of the burst.

\begin{figure}
\begin{center}
\includegraphics[width=0.49\textwidth]{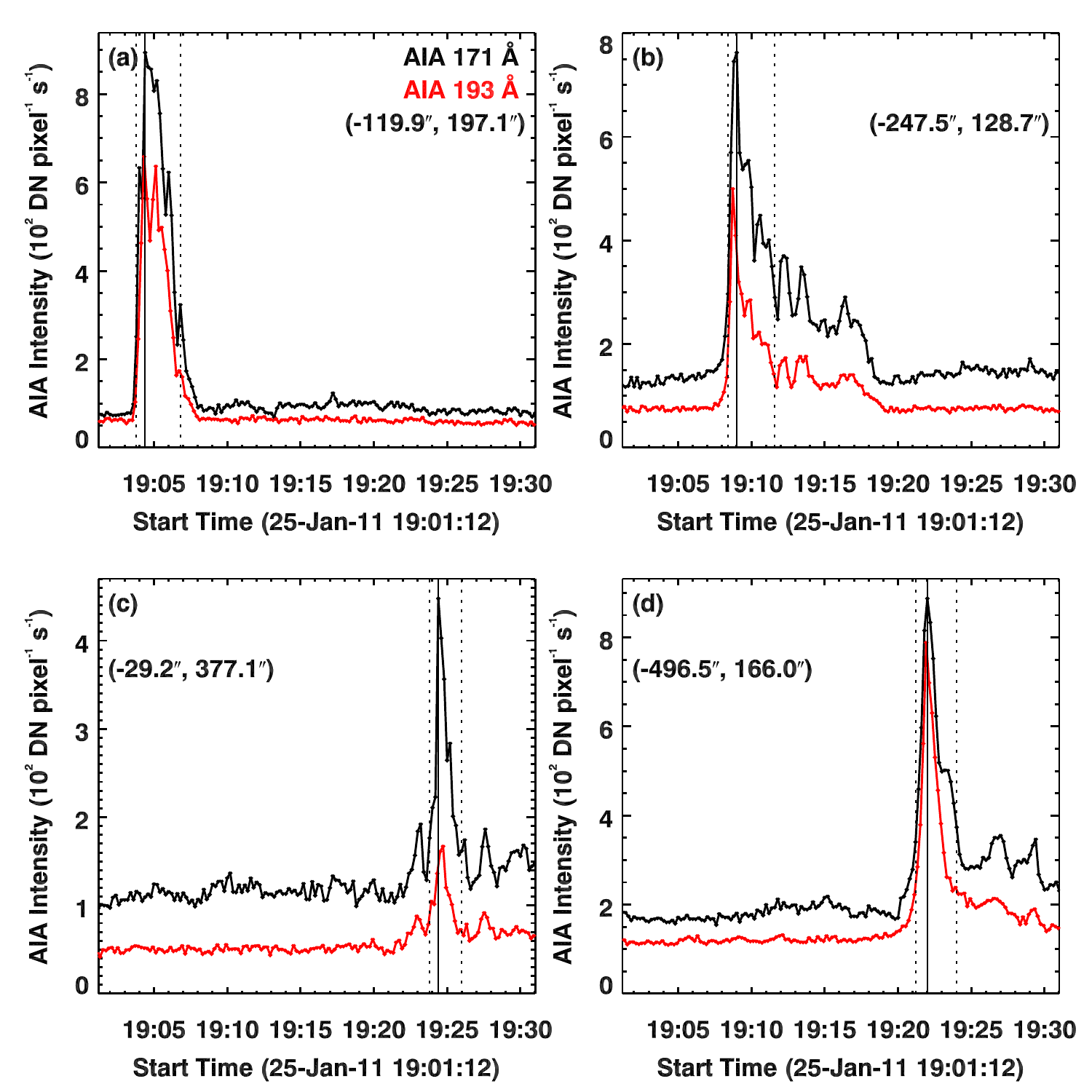}
\caption{Temporal evolution of the quiet-Sun EUV bursts. The time series of four EUV bursts detected in the 2011 dataset are plotted. The light curves show the plasma emission as a function of time at a given spatial pixel over a 30-minute period in that dataset (black shows 171\,\AA, and red represents 193\,\AA). The spatial location of the respective light curve is outlined by a green box in the corresponding panel in Fig.\,\ref{fig:dem1}. The solar $X$ and $Y$ coordinates in arcsec are also quoted. The solid vertical line identifies the peak of the 171\,\AA\ light curve, and the two dotted vertical lines mark the duration of the burst. See Sects.\,\ref{sec:prop} for details.\label{fig:lc1}}
\end{center}
\end{figure}

\subsection{Statistical properties\label{sec:prop}}
We first deduced the pixel-wise lifetime of the burst. From the time step of an identified valid peak, we followed the time series forward and backward in time until before the instance when the intensity first fell below the 1$\sigma$ level above the mean intensity of that time series at that pixel (see illustration in Fig.\,\ref{fig:illus}). In this way, we built a 171\,\AA\ time series that was populated only by bursts over the field of view of $1024\,\rm{pixels}\times 1024\,\rm{pixels}$, that is, the $(614\arcsec)^2$ field of view of interest. A given snapshot therefore includes multiple pixels that capture different stages of the evolution of multiple bursts. The red contours in the top panels of Fig.\,\ref{fig:region} outline pixels with intensities that satisfy our EUV burst criteria. When overlaid on the HMI magnetograms, the bursts are apparently closely clustered in the vicinity of strong magnetic patches (bounded by blue and green contours) that form the solar magnetic network (bottom panels of Fig.\,\ref{fig:region}). This comparison and findings are qualitatively similar to earlier observations: transition region explosive events tend to occur closer to network magnetic fields \citep[e.g.,][]{1994AdSpR..14d..13D}.

\begin{figure}
\begin{center}
\includegraphics[width=0.49\textwidth]{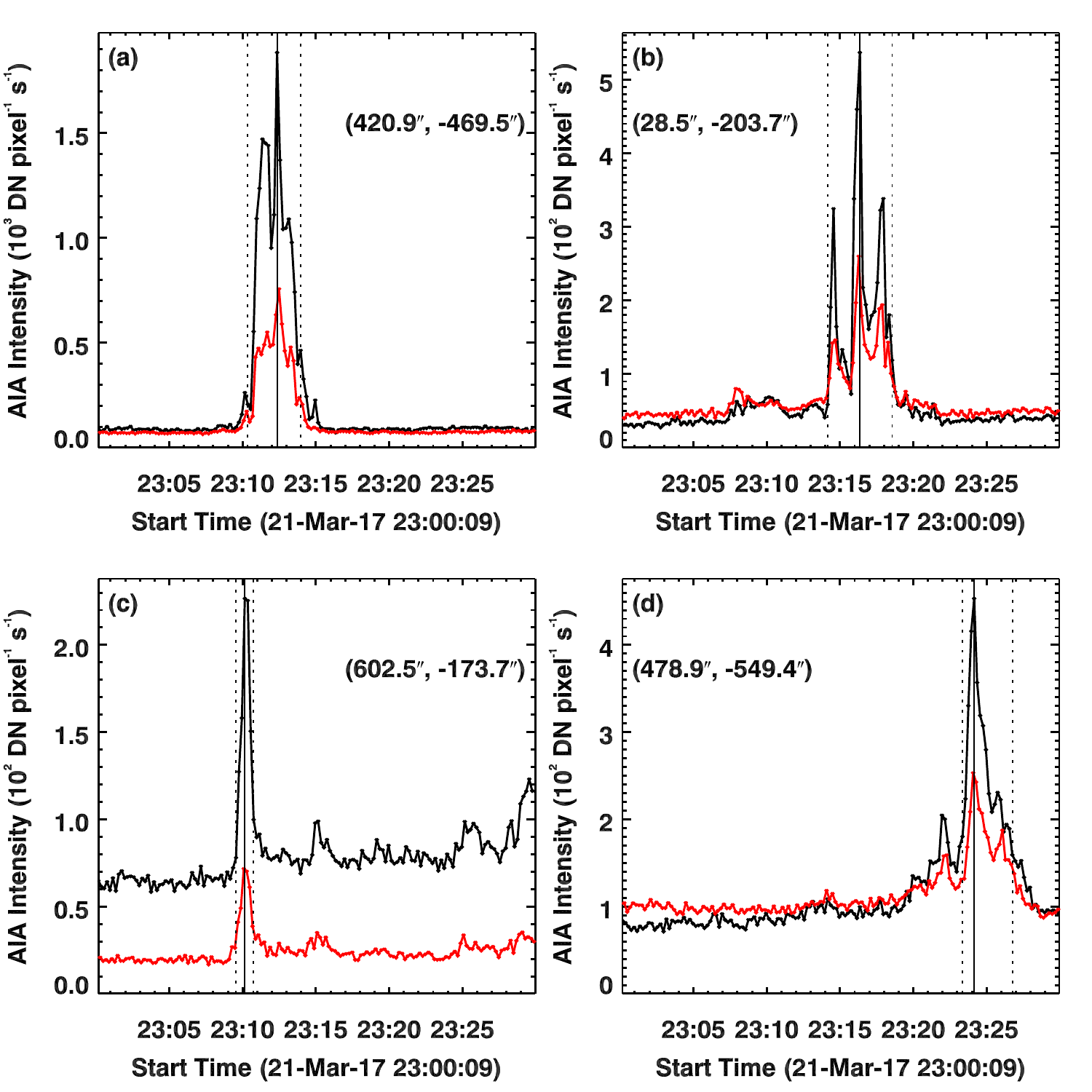}
\caption{Same as Fig.\,\ref{fig:lc1}, but plotted for four bursts detected in the 2017 dataset (see also Fig.\,\ref{fig:dem2}).\label{fig:lc2}}
\end{center}
\end{figure}

These bursts are impulsive in nature. In Figs.\,\ref{fig:lc1} and \ref{fig:lc2} we plot examples of eight bursts that exhibit a variety of impulsive behavior. The duration or lifetime of the burst as identified by applying our method on AIA 171\,\AA\ data is marked with two dotted lines. The time series of these bursts frequently appear to exhibit complex profiles with a superposition of multiple impulsive events, each lasting roughly 60\,s. However, other cases show a simple impulsive profile with a rapid monotonic rise and fall in intensity (e.g., Fig.\,\ref{fig:lc2}c). Nevertheless, a common feature of all the bursts is that they display a sudden intensity enhancement that exceeds the background intensity by several times, and that after the burst, intensity returns to its local pre-event level. Their lifetime is about 100\,s (see below). In addition, as shown in Figs.\,\ref{fig:lc1} and \ref{fig:lc2}, the detected bursts were also observed in AIA 193\,\AA\ filter and typically displayed a similar temporal variability. Compared to the events detected in TRACE observations, which displayed smoother emission curves with intensity variations over timescales of 500\,s\ \citep[Fig.\,6 in][]{2000ApJ...535.1047A}, the bursts we studied are more impulsive.

\begin{figure}
\begin{center}
\includegraphics[width=0.49\textwidth]{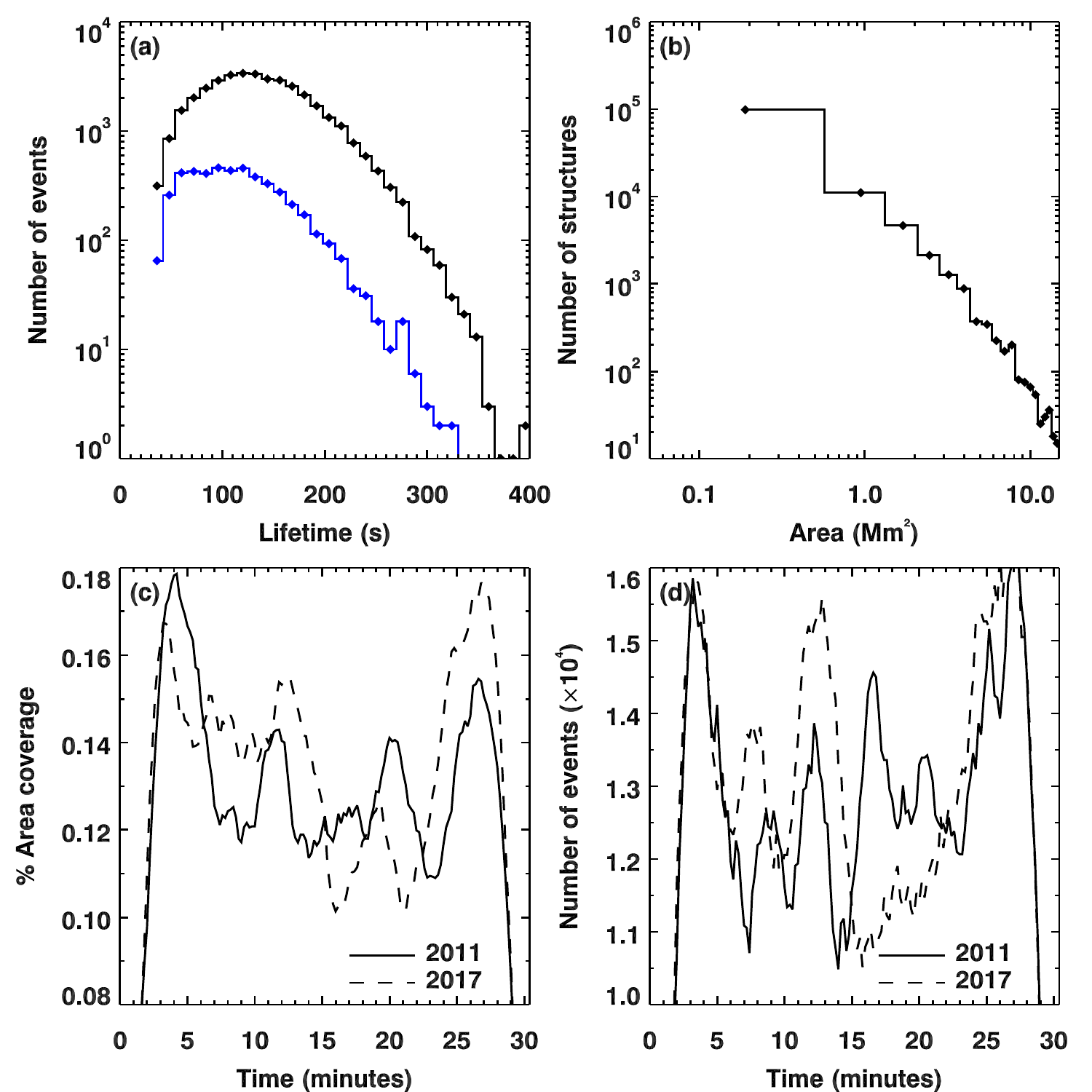}
\caption{Statistical properties of the quiet-Sun EUV bursts observed with the AIA 171\,\AA\ filter. Panels (a) and (b) show the PDF of the lifetime and area of the detected EUV bursts by combining cases from the 2011 and 2017 datasets. In panel (a) the blue histogram shows the PDF of the lifetime of isolated bursts. Panels (c) and (d) display the percentage area coverage of the bursts and the number of events at a given time in the solar corona separately for the two datasets as a function of time. See Sects.\,\ref{sec:prop} for details.\label{fig:props}}
\end{center}
\end{figure}

The probability density function (PDF) of the lifetimes of the bursts deduced with the pixel-wise analysis is shown in Fig.\,\ref{fig:props}(a). By our definition, events with the shortest duration will live for at least 36\,s (the AIA EUV image cadence is 12\,s). Although these short-duration events form a large fraction of the detected events, the peak of the distribution is at a longer timescale of about 120\,s. This indicates that the evolution of the majority of the bursts is already temporally resolved  with the available AIA cadence of 12\,s. As a next step, we counted the number of burst pixels in each snapshot of the time series to compute the area coverage of these events on the Sun (i.e., the area covered by each event was measured as a function of time throughout its lifetime). Because we conducted a pixel-wise analysis, two events separated with a gap of one pixel may likely belong to the same solar structure. To this end, we used the \texttt{morph\_close} function in IDL to fill any gaps or holes of up to one pixel between closely spaced events in every snapshot. These results in terms of the percentage of the area that is covered by the bursts on the Sun are plotted separately for the 2011 (solid) and 2017 (dashed) datasets in Fig.\,\ref{fig:props}c. Both show consistent results: at any given moment in time, the bursts occupy only 0.1\% to 0.15\% of the area on the Sun. 

We then computed the PDF of the area of the burst events. The area PDF shows a power-law-like behavior with an almost linear decrease in the number of events with the increase in area in the log-log plot (Fig.\,\ref{fig:props}b). One key finding is that the peak of the area PDF is at the spatial sampling limit of AIA (i.e., one pixel with a side length of 0.6\arcsec; $1\arcsec\approx 725\,\rm{km}$ near the center of the solar disk). A part of this unresolved population might be explained because the area of an event changes. As we measure the area of an event throughout its lifetime, it is possible that the event first covers one AIA pixel, then grows, and finally decreases again to one pixel. Additionally, some events might cover just one AIA pixel throughout their lifetimes. In either case, this indicates that most of the EUV bursts are unresolved by AIA. That the area PDF does not show a clear peak, unlike the distribution of lifetimes, further suggests that smaller events that are not detected by AIA are more common in the solar corona. Deducing the lifetimes of isolated events that cover just one AIA pixel throughout their evolution could provide information on the optimal spatial resolution and temporal cadence that would be necessary to probe such features. In Fig.\,\ref{fig:props}(a) we plot the PDF of the lifetimes we computed from events that cover only a single AIA pixel (blue histogram). The lifetime PDF of these unresolved events is very similar to that of the complete sample, suggesting that bursts might show a similar temporal evolution even at smaller spatial scales.

Furthermore, we also plot the number of events (after filling the gaps) in a given snapshot as a function of time in Fig.\,\ref{fig:props}(d). Based on our analysis, we estimate that there are about 12000 to 13000 EUV bursts at some stage in their evolution in the quiet-Sun corona at any given instance. Similar to the time series of the area coverage, the number of structures we estimate in both datasets is consistent. When we assume a burst lifetime of 120\,s, the event birth rate is about 100\,s$^{-1}$. This is on the same order of magnitude as the birthrate of transition region explosive events \citep[][]{1994AdSpR..14d..13D,2003A&A...409..755H}.

Compared to the quiet-Sun corona, EUV emission is fainter than coronal hole regions. This lower emission generally results from the presence of large-scale open magnetic structures that allow plasma to easily expand outward. In the 2017 dataset, a significant area of the analyzed field of view is covered by a coronal hole (northwest quadrant of the top right panel in Fig.\,\ref{fig:region}). However, the analyzed EUV bursts are uniformly distributed throughout the field of view, and there is no clear distinction between quiet-Sun and coronal holes in this regard. Although not shown, we found no difference between the lifetime distributions of bursts in either dataset, likely because the analyzed events are more local and are probably not affected by the large-scale open magnetic fields in the coronal holes. Furthermore, the number of events and the area coverage we present in Fig.\,\ref{fig:props} from the 2011 and 2017 datasets is comparable. We therefore suggest that quiet-Sun and coronal holes are not clearly distinct with respect to EUV bursts. We also obtained statistically similar results for bursts detected in the AIA 193\,\AA\ and 211\,\AA\ filters (see Appendix\,\ref{sec:app}).

\subsection{Emission characteristics\label{sec:emiss}}
\begin{figure*}
\begin{center}
\includegraphics[width=0.75\textwidth]{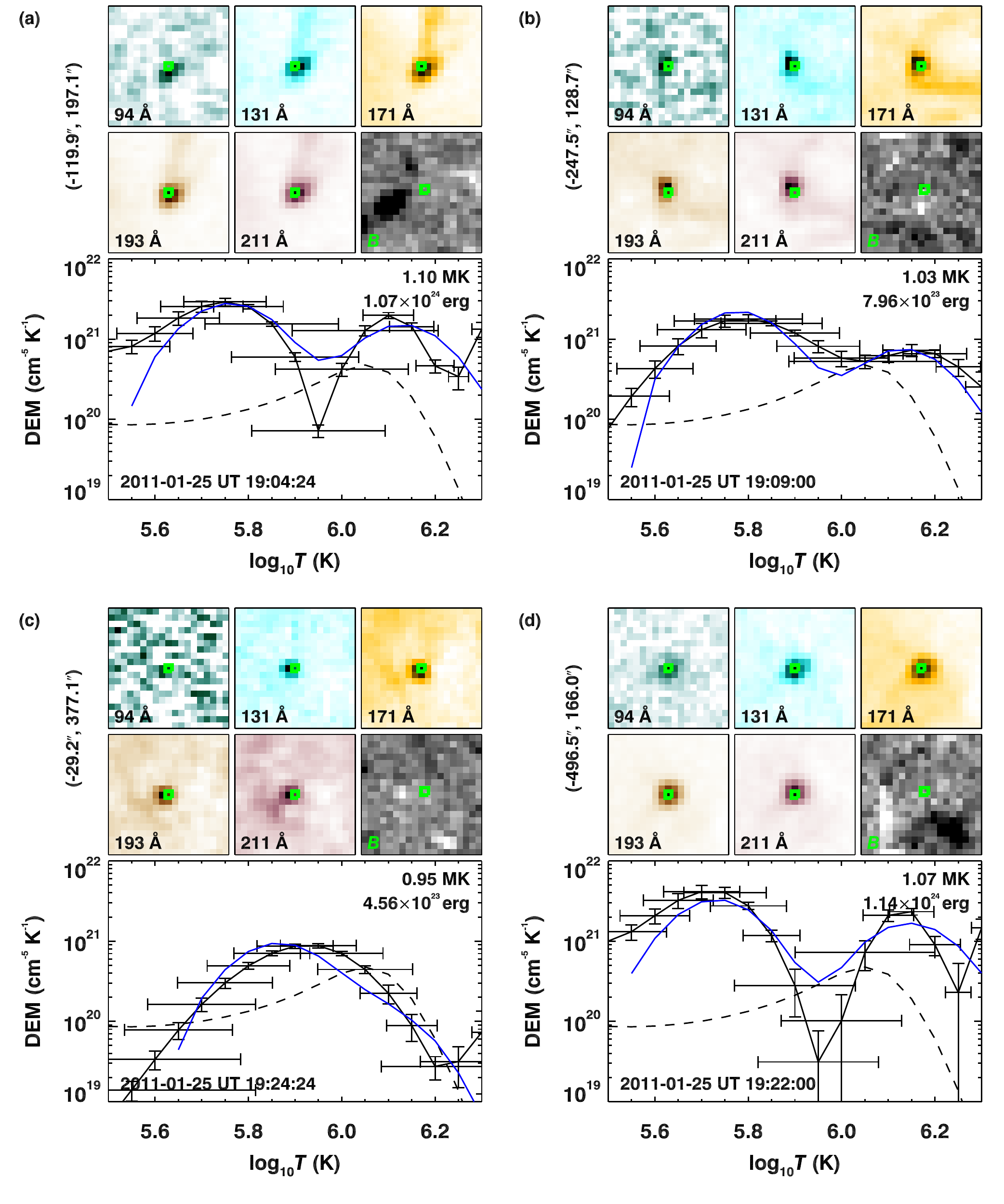}
\caption{Spatial morphology and emission measure distribution of the selected EUV bursts. The top segment of each panel shows maps covering the burst in five of the six AIA EUV filters (in inverted color), and the lower right tile presents the cotemporal HMI line-of-sight magnetic field map, saturated at $\pm$30\,G. The field of view is about $11.4\arcsec\times 11.4\arcsec$. The AIA 171\,\AA\ and 193\,\AA\ time series from the green box regions are plotted in the respective panels in Fig.\,\ref{fig:lc1}(a)-(d). The spatial coordinates of the green box are listed at the left side of the top segment. The bottom segment in each panel is a plot of the DEM as a function of temperature derived using a regularized inversion technique \citep[solid black; method-I;][]{2012A&A...539A.146H} and a sparse inversion technique \citep[solid blue; method-II;][]{2015ApJ...807..143C} by combining the AIA EUV emission in the green box regions from the six filters. The vertical black bars are the 1$\sigma$ errors in DEMs, and the horizontal black bars are associated with the energy resolution of the regularized inversion technique. The dashed curve is the quiet-Sun DEM profile based on Hinode observations \citep[][]{2009ApJ...705.1522B}, available in the CHIANTI atomic database \citep{1997A&AS..125..149D,2019ApJS..241...22D}. The emission-weighted temperature of the plasma (in MK) and thermal energy content of the burst (in erg) are quoted. The time stamp corresponds to the snapshots in the top segment (same as the peak time identified with the vertical line in Fig.\,\ref{fig:lc1}). These bursts are detected in the 2011 dataset. See Sects.\,\ref{sec:emiss} for details.\label{fig:dem1}}
\end{center}
\end{figure*}

\begin{figure*}
\begin{center}
\includegraphics[width=0.75\textwidth]{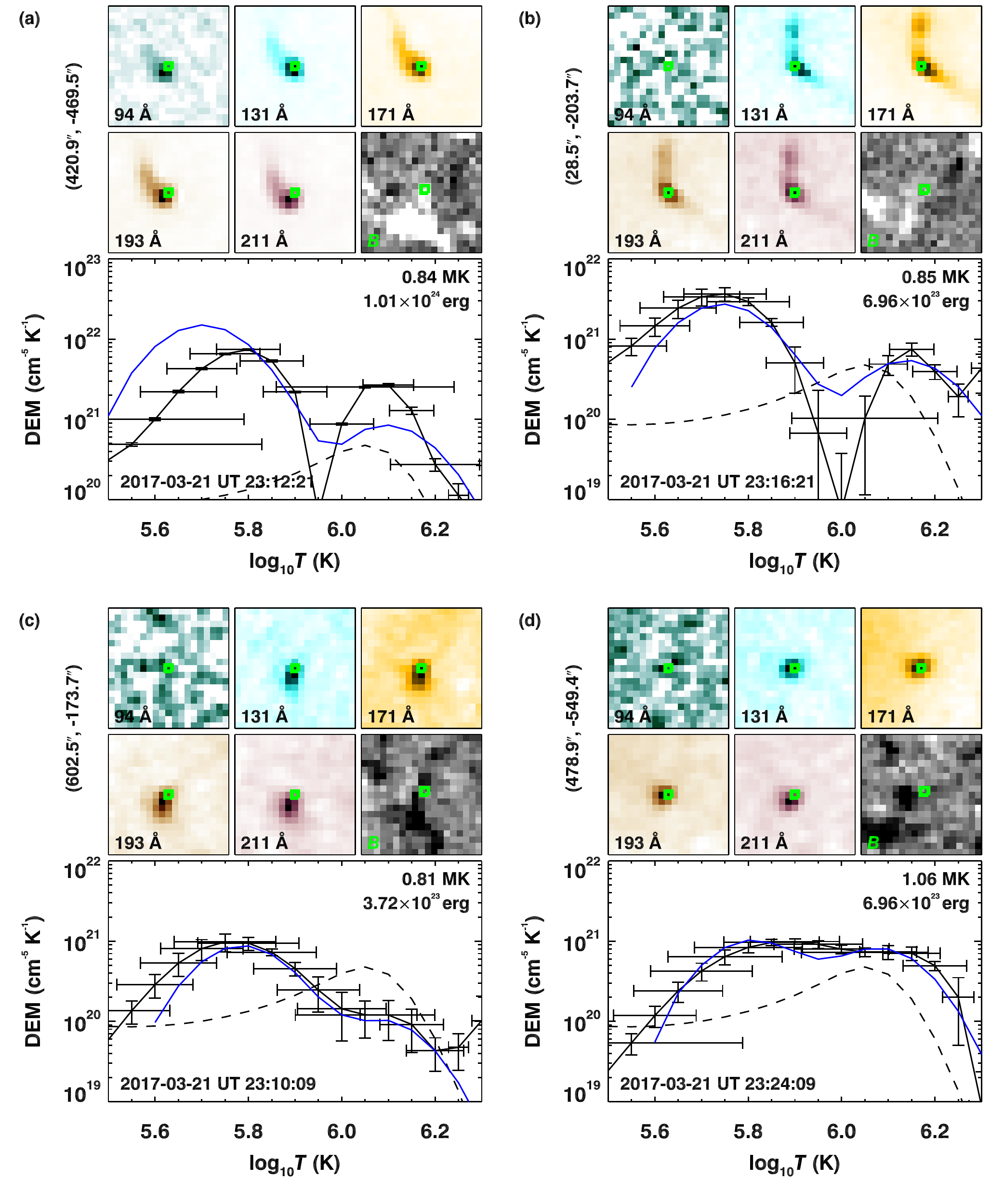}
\caption{Same as Fig.\,\ref{fig:dem1}, but plotted for the four bursts in the 2017 dataset. The AIA 171\,\AA\ and 193\,\AA\ time series from the green box regions are plotted in the corresponding panels in Fig.\,\ref{fig:lc2}(a)-(d). See Sects.\,\ref{sec:emiss} for details. \label{fig:dem2}}
\end{center}
\end{figure*}

Spatially resolved EUV bursts (i.e., those that span more than two AIA pixels) appear as compact blobs in multiple AIA filters (top segments in Figs.\,\ref{fig:dem1} and \ref{fig:dem2}). Some of them also show flame-like extension connected to a bright central core (e.g., Fig.\,\ref{fig:dem1}(a) and Figs.\,\ref{fig:dem2}(a)-(b)). Their spatial structuring is similar to that of the IRIS UV bursts that have been observed in solar active regions \citep[][]{2014Sci...346C.315P,2018SSRv..214..120Y}. However, the main difference is that while the IRIS active region UV bursts are rarely seen in the EUV images\footnote{A few observations of active region UV bursts and quiet-Sun explosive events have been reported in both IRIS and multiple AIA EUV filter diagnostics \citep[e.g.,][]{2014ApJ...797...88H,2015ApJ...809...82G,2018ApJ...856..127G}.}, the bursts studied here (by definition) do appear in AIA diagnostics.

To deduce the thermal characteristics of these bursts, we calculated the differential emission measure (DEM) at that spatial location and time of the AIA 171\,\AA\ peak (the location is marked with green boxes in the upper segments of panels a and b of Figs.\,\ref{fig:dem1} and \ref{fig:dem2}). If these EUV bursts are related to explosive events, plasma at transition region temperatures below 1\,MK will contribute to the emission. If present, emission from cooler plasma will be recorded by the AIA 131\,\AA\ filter, which has its (lower temperature) response peak around log$_{10}T \rm{(K)}$ of 5.5. Although the bursts may also emit at even lower temperatures, the response of the AIA filters drops significantly below log$_{10}T \rm{(K)}$ of 5.5. At the higher end, the bursts may be heated to quiet-Sun coronal temperatures of 1 to 2\,MK. The DEMs are therefore computed over the temperature range of log$_{10}T \rm{(K)}$ 5.5\textendash6.3 using emission from all the six AIA EUV channels at the time when the AIA 171\,\AA\ emission peaked.

We employed two independent methods to calculate to DEMs, and they essentially yield the same results: a regularized inversion technique \citep[method-I;][]{2012A&A...539A.146H}, and a sparse inversion technique \citep[method-II;][]{2015ApJ...807..143C}. The input parameters passed to these methods are presented in Appendix\,\ref{sec:dpar}. The results are displayed in the bottom segments of Figs.\,\ref{fig:dem1} and \ref{fig:dem2}. In six out of the eight cases, the DEMs show a double-peaked structure with a stronger contribution to the emission from temperatures around log$_{10}T \rm{(K)}$ 5.6\textendash5.8, corresponding to the transition region (solid curves). The secondary component peaks around log$_{10}T \rm{(K)}$ 6.1. In two cases (Fig.\,\ref{fig:dem1}c and Fig.\,\ref{fig:dem2}d), the DEM distribution is consistent with only one broad component. We compared these burst DEMs with the DEM derived by \citet{2009ApJ...705.1522B} of a quiet-Sun region using Hinode observations (dashed curves in Figs.\,\ref{fig:dem1} and \ref{fig:dem2}). The Hinode quiet-Sun DEM has a single peak above 1\,MK with a broad tail toward the lower temperatures and a steep negative slope toward higher temperatures, which is also the case for some quiet-Sun coronal bright points \citep[e.g.,][]{2013ApJ...768...32C}. In comparison, the bursts show a clear enhancement in the emission from lower temperatures below 1\,MK. At higher temperatures above 1\,MK, however, the DEM shapes vary. In particular, except for two cases (Fig.\,\ref{fig:dem1}c and Fig.\,\ref{fig:dem2}c), there is a clear difference between the DEM shapes above 1\,MK in the remaining examples. Either the DEM peaks above 1\,MK are at slightly higher temperatures than the quiet-Sun DEM, and/or the magnitude of the DEM is larger at higher temperatures. This tentatively indicates the presence of hotter coronal plasma (i.e., $>$1\,MK) in at least some of the bursts. Overall, our DEM analysis shows that the bursts have a broad temperature distribution. 

\begin{figure}
\begin{center}
\includegraphics[width=0.49\textwidth]{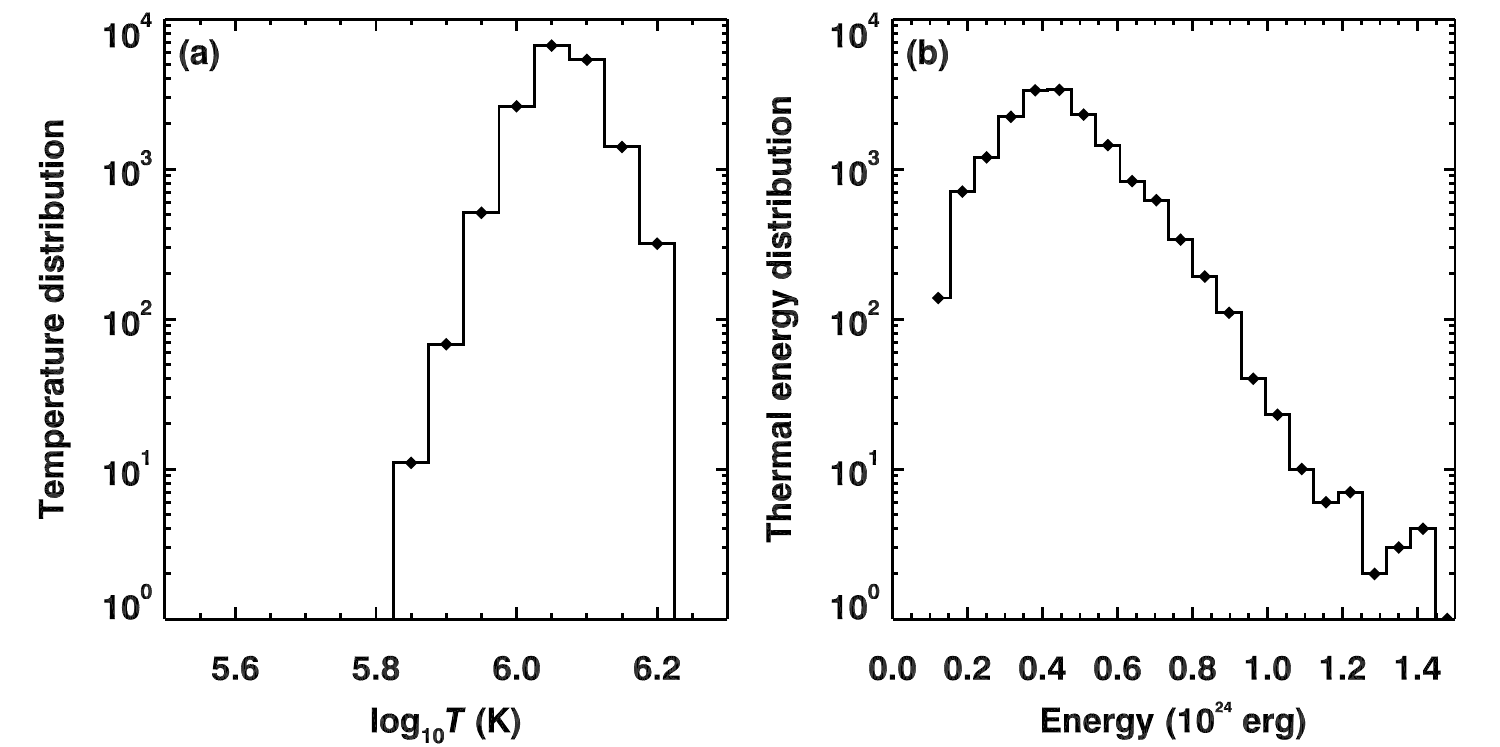}
\caption{Temperature and thermal energy distributions of EUV bursts. PDFs of emission-weighted temperature (panel a) and thermal energy content (panel b) derived at the peak of all detected bursts in the 2011 and 2017 datasets are plotted.\label{fig:eth}}
\end{center}
\end{figure}

We can estimate the thermal energy content of the burst through the volumetric emission measure, $EM$. This is defined through $EM=\int n^2\,{\rm{d}}V \approx f\,n^2\,V$, where $n$ is the electron number density, $V$ is the volume under consideration, and $f$ is the filling factor of the emitting volume. The thermal energy content is given through $E_{\rm{th}}=3\,N\,k_{\rm{B}}\,T$, where $k_{\rm{B}}$ is the Boltzmann constant, $T$ is the plasma temperature, and $N=nV$ is the total number of electrons in the volume. With the emission measure we can now express the thermal energy content as $E_{\rm{th}}=3\,k_{\rm{B}}\,T (EM{\cdot}V/f)^{1/2}$. The plasma temperature is assumed to be equal to the emission-weighted temperature, and the $EM$ is obtained by integrating the DEM over temperature over an area covered by one AIA pixel. The volume, $V$, of the emitting plasma is assumed to be a cube with a side length of one AIA pixel. Because the displayed bursts clearly extend over one AIA pixel, we further assumed that $f=1$. The emission-weighted temperatures of these eight bursts is in the range of 0.8\,MK to 1\,MK, and the resulting thermal energy is in the range of $10^{23}$\,erg to 10$^{24}$\,erg (Figs.\,\ref{fig:dem1} and \ref{fig:dem2}). For all of the detected events in the 2011 and 2017 datasets, we calculated the emission-weighted temperature and the thermal energy (derived from DEMs based on method-I) of each burst at its peak assuming $f=1$. The respective PDFs are plotted in Fig.\,\ref{fig:eth}. The emission-weighted temperature spans from 0.6\,MK to 1.6\,MK with a clear peak around 1\,MK. The thermal energy of the bursts is in the range of $10^{23}$\,erg to 10$^{24}$\,erg, with a peak around $4\times 10^{23}$\,erg.\footnote{Similar to other statistical properties, we found no systematic difference between the distribution of the energy content of bursts in the 2011 and 2017 datasets.} This energy content is only a lower limit for the total energy put into the burst because during the heating, energy will be lost, for example, through heat conduction and other processes. Overall, these energies are comparable to the energy of typical nanoflares \citep[][]{1988ApJ...330..474P}. Based on the statistical properties presented in Fig.\,\ref{fig:props} and the thermal energy analysis in Fig.\,\ref{fig:eth}, we define EUV bursts as compact nanoflare events covering an area smaller than 10\,Mm$^2$ with lifetimes of about 100\,s, and with an energy content of about 10$^{24}$\,erg. We exclude any persistently bright features (e.g., coronal bright points) in our definition of EUV bursts.

\section{Discussion\label{sec:disc}}

\begin{figure}
\begin{center}
\includegraphics[width=0.49\textwidth]{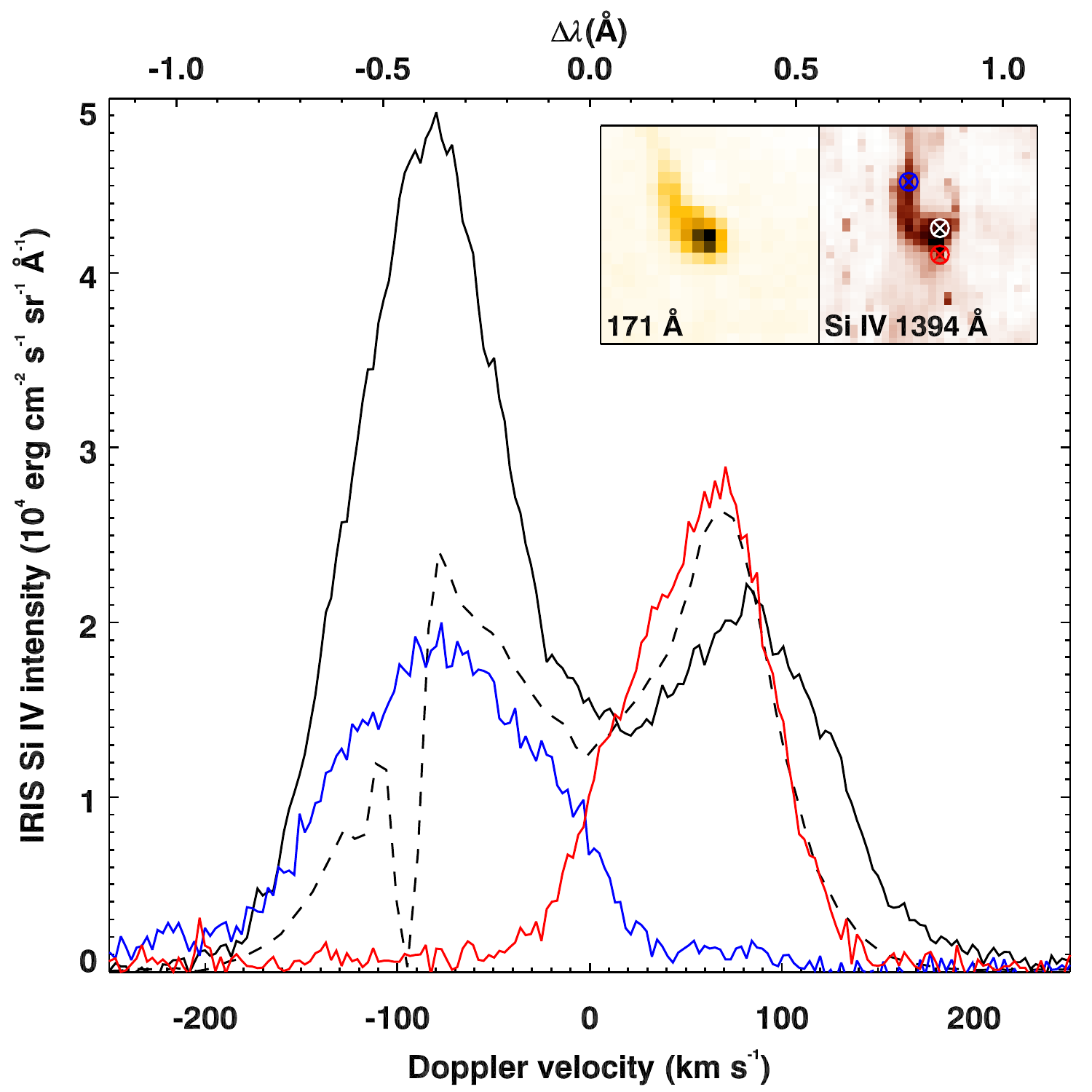}
\caption{Spectral profiles emergent from a burst. The solid black, blue, and red curves represent Si\,{\sc iv} 1394\,\AA\ spectral profiles as a function of Doppler velocity, recorded by IRIS at the site of an EUV burst observed with SDO/AIA. Here $0\,\rm{km}\,\rm{s}^{-1}$\ corresponds to the rest wavelength of 1393.755\,\AA. The associated maps of the EUV burst are displayed in the inset. The left tile in the inset is the same as in Fig.\,\ref{fig:dem2}a. The right tile in the inset is the Si\,{\sc iv} 1394\,\AA\ raster map in inverted colors (integrated over $\pm200\,\rm{km}\,\rm{s}^{-1}$), covering the EUV burst. The white, blue, and red symbols mark the position of the spatial pixels we used to extract the black, blue, and red spectral profiles, respectively. For comparison, the dashed black curve is the Si\,{\sc iv} 1394\,\AA\ spectral profile emergent from an active region UV-burst (labeled 1) discussed in \citet{2014Sci...346C.315P}. The profile is scaled by a factor of 0.02. The strong dip in the dashed Si\,{\sc iv} profile around $-100\,\rm{km}\,\rm{s}^{-1}$ is the Ni\,{\sc ii} absorption profile. See Sects.\,\ref{sec:disc} for details.\label{fig:spect}}
\end{center}
\end{figure}

Observations with AIA do not (directly) contain information of the plasma flows. Therefore we employed spectra taken with IRIS during the time of the dataset from 2017. The burst in Fig.\,\ref{fig:dem2}a is captured with the IRIS slit and thus offers a probe into plasma flows at the time of the event. In Fig.\,\ref{fig:spect} we show IRIS Si\,{\sc iv} spectral profiles from three regions, one at the core (solid black curve), and a pair from the two sides of the core (blue and red curves). The spectral profile from the central core region displays two peaks at $\pm$90\,km\,s$^{-1}$, consistent with bidirectional plasma jets produced at the site of magnetic reconnection \citep[][]{1997Natur.386..811I,2014Sci...346C.315P}. At one side of the core in the extended flame-like segment, IRIS recorded a $-90$\,km\,s$^{-1}$ blueshifted Si\,{\sc iv} spectral profile, suggesting plasma upflow. At the other side, the spectral profile is redshifted by $+90$\,km\,s$^{-1}$, indicative of plasma downflows. In comparison to an active region UV burst discussed in \citet{2014Sci...346C.315P} (dashed curve in Fig.\,\ref{fig:spect}), the EUV burst in this case radiated an intensity that was an order of magnitude lower. However, the speed of the bidirectional jets in both cases is almost similar and comparable to the outflow speeds in transition region explosive events \citep[][]{1994AdSpR..14d..13D}.

Observations and numerical models show that chromospheric and transition region UV bursts in active regions are triggered when patches of opposite magnetic polarities connected through a U-type field configuration cancel as they approach each other at the solar surface \citep[e.g.,][]{2014Sci...346C.315P,2018SSRv..214..120Y,2019A&A...626A..33H}, similar to photospheric Ellerman bombs \citep[][]{2002ApJ...575..506G}. Bursts like this are also triggered in a fan-spine magnetic configuration created by patches of minor magnetic polarity embedded in regions with stronger magnetic fields of opposite polarity \citep[e.g.,][]{2017A&A...605A..49C,2019A&A...628A...8P}. Nevertheless, a key aspect in both these configurations is the release of magnetic energy through interaction of opposite-polarity magnetic fields \citep[e.g.,][]{2018ApJ...862L..24P}. However, although the signature of magnetic reconnection as evident from bidirectional jets is clear, the EUV burst in Fig.\,\ref{fig:spect} directly overlies an apparent unipolar magnetic field region of positive polarity with no clear signatures of the adjacent interacting negative polarity magnetic field (see the HMI map in Fig.\,\ref{fig:dem2}a). It is possible that a small-scale weak opposite-polarity magnetic element is present, but remained undetected in HMI observations as a result of a combination of the moderate spatial resolution and sensitivity of the instrument \citep[e.g.,][]{2017ApJS..229....4C,2019A&A...623A.176C}. It is also possible that the burst is triggered by a component magnetic reconnection within the unipolar magnetic patch. 

In addition, the DEMs are consistent with a broad temperature distribution in the bursts with contribution from transition region plasma (below 1\,MK) to a large part and to a lesser degree, likely also some from coronal plasma (above 1\,MK). Spectroscopic observations are necessary to determine the thermal characteristics of these bursts. IRIS samples plasma mostly at or below 0.1\,MK\ and is therefore not suitable for coronal diagnostics, although \citet{2019ApJ...871...82G} detected Fe\,{\sc xii} emission forming at coronal temperatures with IRIS in a singular case of an active region UV burst that was more energetic (due to stronger magnetic fields) than the bursts studied here. Using observations from SUMER on board SOHO, \citet{2002A&A...392..309T} previously found no coronal counterparts of explosive events. Their sample was limited to only three cases, however. In principle, the EUV imaging spectrometer (EIS) on Hinode could be used for such a study, in particular because of its high sensitivity in Fe\,{\sc xii} near 195\,\AA. However, the spatial resolution of EIS is significantly lower than that of AIA, and thus EIS would only capture the (fewer) larger events. A future study should be conducted to determine how EIS can help to settle this problem.

A remaining key question is the role of these bursts in quiet-Sun coronal heating. Through the statistics of bursts presented in Fig.\,\ref{fig:props}, that is, the area coverage and number of events, it is clear that these events are not pervasive enough to account for energy losses from the quiet-Sun corona as a whole, even when we assume that some of them do reach million Kelvin temperatures. The energy loss from the quiet-Sun corona is about $10^{28}$\,erg\,s$^{-1}$ \citep[][]{1977ARA&A..15..363W}. Typically, the energy content of these bursts will follow a power-law-like distribution \citep[e.g.,][]{2000ApJ...535.1047A}. To obtain a simple estimate of the number of such events that are required to balance the quiet-Sun coronal energy losses, we assume that all the bursts detected here are nanoflares, with a total energy output of $10^{24}$\,erg during their lifetime (see Figs.\,\ref{fig:dem1}, \ref{fig:dem2}, and \ref{fig:eth}). Based on the peak of the lifetime distribution (Fig.\,\ref{fig:props}a), the bursts supply energy to the quiet-Sun\ corona at a rate of about $10^{22}$\,erg\,s$^{-1}$. This means that at any given moment in time, there should be about $10^6$ such events on the Sun. From Fig.\,\ref{fig:props}d, we estimate that there are only $10^4$ events at any moment in time on the Sun. The number of detected events is therefore at least 100 times lower than what is required to support coronal energy requirements.  

Although the energy content of many of the bursts we see is comparable to that of nanoflares, they are not spatially resolved in the AIA observations. Here we can expect significant progress from the recently launched Solar Orbiter mission \citep[][]{2020A&A...642A...1M}. In the context of our study, the Extreme Ultraviolet Imager \citep[EUI;][]{2020A&A...642A...8R}, the Polarimetric and Helioseismic Imager \citep[PHI;][]{2020A&A...642A..11S}, and the SPectral Imaging of the Coronal Environment \citep[SPICE;][]{2020A&A...642A..14S} are of particular interest. During the nominal mission phase, the first perihelion in March 2022 will bring the spacecraft as close as about 0.3 astronomical units. At this distance, the spatial resolution of EUI will correspond to about 0.3{\arcsec} as seen from Earth orbit. With this it will provide images with the highest resolution that were ever taken of the solar corona, significantly better than AIA, and slightly better than the data from the two 5-minute rocket flights of the High-Resolution Coronal Imager \cite[Hi-C;][]{2019SoPh..294..174R}. With matching spatial resolution, the measurements of the magnetic field vector in the photosphere by PHI will provide information on the magnetic driver of the bursts. With its coverage of spectral lines from 10$^4$\,K to well above 10$^6$\,K, SPICE will provide the thermal structure of these events, but at significantly lower spatial resolution.

The remote-sensing instruments of Solar Orbiter have so far provided only commissioning data. Some of these are of high quality, however, as announced in an ESA press release\footnote{\tiny\url{https://sci.esa.int/web/solar-orbiter/-/solar-orbiter-s-first-images-reveal-campfires-on-the-sun}}. In particular, the so-called campfires found by EUI \citep{bergh2021} might be related to the events we investigated in our study. We wonder whether these campfires are similar in nature to the events we studied here. Clearly, both are brightenings in the EUV near 171\,{\AA} (one channel of EUI is centered around 174\,{\AA}). It would be interesting to speculate if their nature is similar to the reconnection driven jets attributed to the transition region explosive events \citep[][]{1994AdSpR..14d..13D,1997Natur.386..811I}, but now extended to higher temperatures. This would at least be suggested by our analysis of the emission measure that is enhanced at coronal temperatures in most of the events we studied (see Figs.\,\ref{fig:dem1} and \ref{fig:dem2} and Sect.\,\ref{sec:emiss}). Future co-observations between EUI and AIA will allow for a detailed comparison with the campfires reported by EUI.

\section{Conclusions\label{sec:conclusions}}

Based on our estimates, we conclude that these bursts are not sufficient to power the corona. This is broadly consistent with earlier results \citep[e.g.,][]{2000ApJ...535.1047A,2016A&A...591A.148J}, although we used AIA data with a higher cadence for the event detection. The quiet-Sun coronal heating may not burst-like, and perhaps it is largely driven by magnetohydrodynamic waves \citep[e.g.,][]{2011Natur.475..477M}. We did not investigate whether and how bursts might affect coronal features away from the central region, that is, the detected event. The question remains whether they also power connected loops in a quasi-steady manner, or if they might drive waves. For example, in Fig.\,\ref{fig:region}, the question also remains whether the bursts that are detected at the footpoint regions of coronal bright points play any role in powering the overlying loops. This is very similar to the existence of reconnection-driven UV bursts at the footpoints of hotter loop systems over 5\,MK\ in the cores of active regions \citep[][]{2018A&A...615L...9C,2020A&A...644A.130C}, but at lower temperatures of 1\,MK. It remains to be seen whether EUI detects more if not all of the missing bursts (or in the terminology of the ESA press release, the campfires). The combination of EUI, SPICE, and PHI on Solar Orbiter will provide us with new opportunities to address the nature of the bursts seen in the quiet Sun and their role in heating the corona in general.

\begin{acknowledgements}
We thank the anonymous referee for providing constructive comments that helped us improve the manuscript. PRY acknowledges funding from the NASA Heliophysics Guest Investigator Program. SDO data are courtesy of NASA/SDO and the AIA, EVE, and HMI science teams. IRIS is a NASA small explorer mission developed and operated by LMSAL with mission operations executed at NASA Ames Research Center and major contributions to downlink communications funded by ESA and the Norwegian Space Centre. CHIANTI is a collaborative project involving George Mason University, the University of Michigan (USA), University of Cambridge (UK) and NASA Goddard Space Flight Center (USA). This research has made use of NASA’s Astrophysics Data System.
\end{acknowledgements}

\begin{appendix}
\section{Statistics of EUV bursts detected in the AIA 193\,\AA\ and 211\,\AA\ filters\label{sec:app}}
\begin{figure*}
\begin{center}
\includegraphics[width=0.49\textwidth]{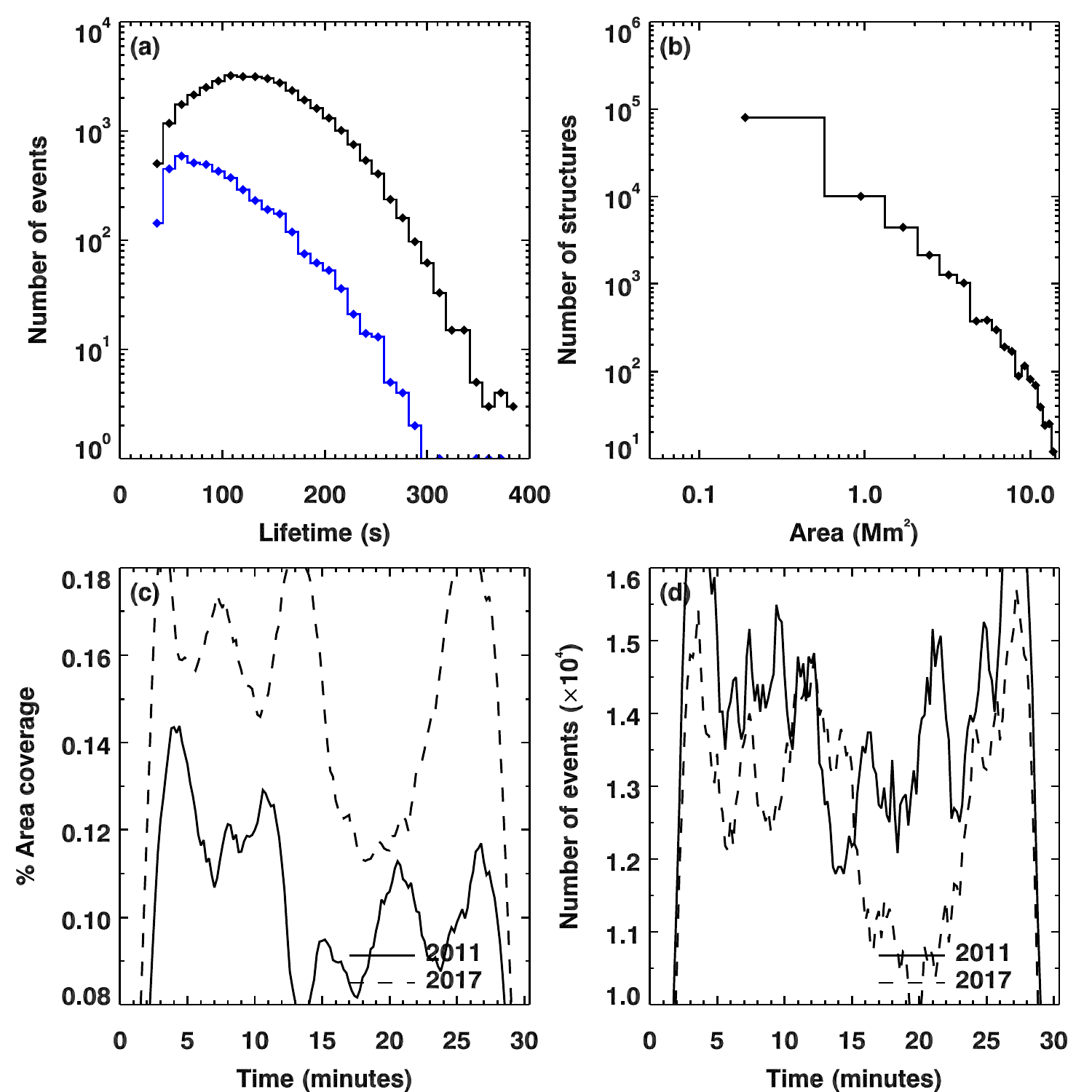}
\includegraphics[width=0.49\textwidth]{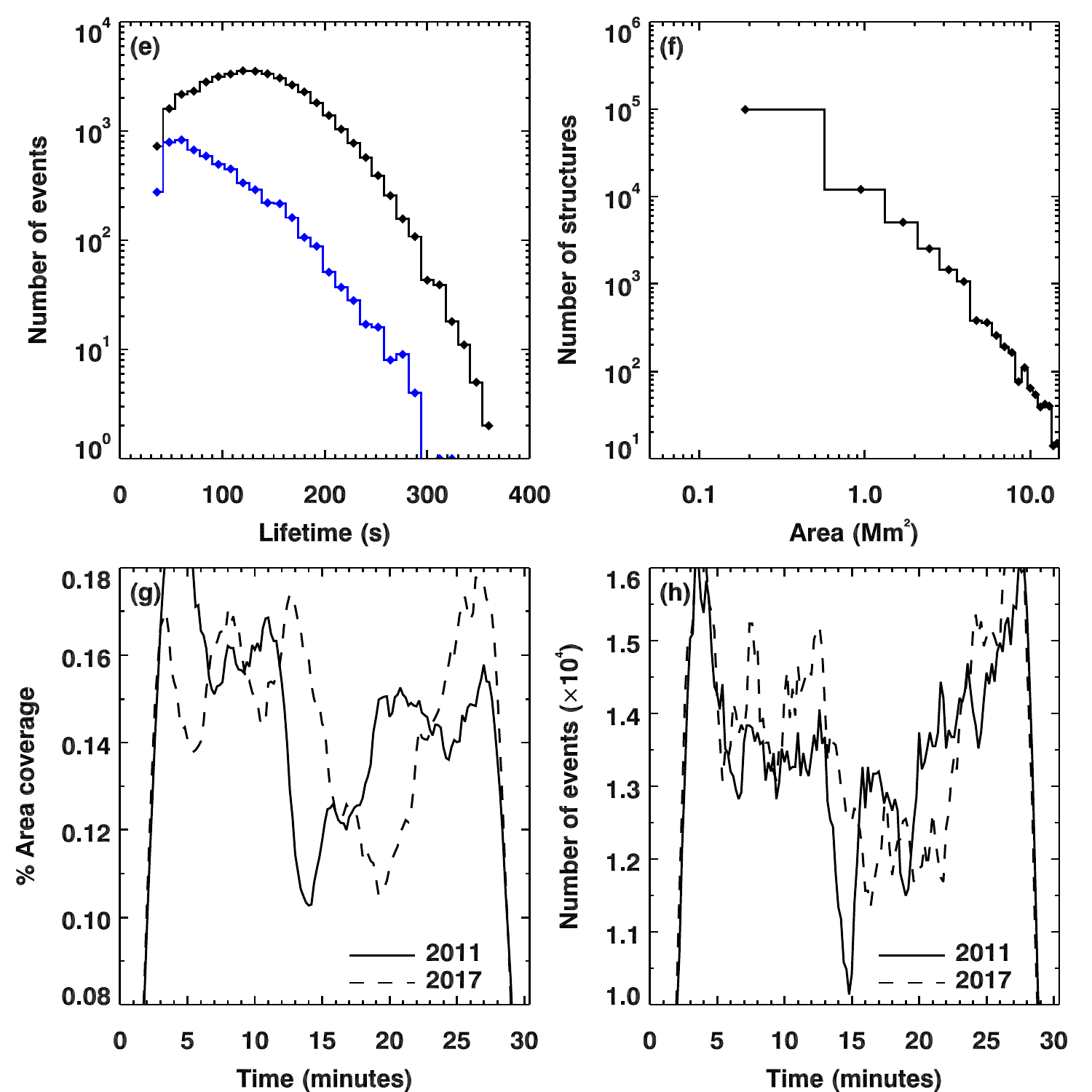}
\caption{Statistical properties of the quiet-Sun EUV bursts observed with the AIA 193\,\AA\ and 211\,\AA\ filters. Same as Fig.\,\ref{fig:props}, but plotted for bursts detected in the AIA 193\,\AA\ filter (panels a-d) and in the AIA 211\,\AA\ filter (panels e-h). In panel (c), the solid line is multiplied by 1.5. To facilitate comparison of the statistics in different filters, the scales of the vertical axes in the respective panels are fixed to be the same as in Fig.\,\ref{fig:props}. See Sect.\,\ref{sec:burst} and Appendix\,\ref{sec:app} for details.\label{fig:props2}}
\end{center}
\end{figure*}

In the main text, we discussed various properties of the EUV bursts detected in the AIA 171\,\AA\ filter. However, some bursts may be detected with filters that respond to hotter plasma (e.g., AIA 193\,\AA\ and 211\,\AA\ filters) and are not present in the AIA 171\,\AA\ filter, which responds to emission from relatively cooler plasma and vice versa. To this end, we also applied our burst detection algorithm (Sect.\,\ref{sec:burst}) to the AIA 193\,\AA\ and 211\,\AA\ filters and derived their statistical properties (lifetime, area coverage, and number of events). These results are displayed in Fig.\,\ref{fig:props2}. These properties are generally consistent in the three filters (see also Fig.\,\ref{fig:props}). This suggests that these bursts are all captured by the three filters, whose responses overlap around 1\,MK,
which is also consistent with our DEM analysis, which shows a peak of emission-weighted temperature at 1\,MK. In principle, a similar analysis might be applied to the other three AIA EUV filters, which are centered around 94\,\AA, 131\,\AA, and 335\,\AA. However, their response in quiet-Sun regions is dominated by noise, therefore they are not suitable for such an analysis.

\section{Input parameters to DEM analysis\label{sec:dpar}}
We used two methods to calculate DEMs: a regularized inversion technique \citep[method-I;][]{2012A&A...539A.146H} and a sparse inversion technique \citep[method-II;][]{2015ApJ...807..143C}. Here we briefly describe the input parameters that were passed to these methods to compute DEMs. For both methods, the time-dependent AIA response functions are calculated by setting the keywords \texttt{evenorm}, \texttt{chiantifix}, and \texttt{noblend} to 1. The uncertainties in AIA EUV intensities are calculated using the \texttt{aia\_bp\_estimate\_error} procedure available in IDL/solarsoft.

In method-I, we used the regularization tweak parameter set to 1, and the regularization multiplying factor set to 1.5. No initial guess solution is used. The AIA uncertainties are multiplied by a factor 1.2 to aid the method to arrive at a positive solution. Furthermore, internally, the method uses the regularization multiplying factor to increase the tweak parameter in each iteration and tries to arrive at a positive solution.

In method-II, we used a tolerance factor set to 1.2. Internally, this factor is used by the method as a multiplication factor to the AIA uncertainties. This tolerance factor is adaptively increased by the method to arrive at a positive solution. The basis functions are Dirac-delta functions plus Gaussians with three bases sigmas (as defined in the method) of 0.125 in log$_{10}T$.

\end{appendix}
\end{document}